\documentclass{JUSTC}

\usepackage{algorithm,algorithmicx,algpseudocode}%
\usepackage{enumitem}

\DeclareMathOperator*{\cov}{cov}

\newtheorem{condition}{Condition}
\newtheorem{lemma}{Lemma}
\newtheorem{assumption}{Assumption}
\newtheorem{corollary}{Corollary}

\type{Manuscript type}
\title{Online Kernel Sliced Inverse Regression}

\author[1,\Letter]{Wenquan Cui}

\author[1]{Yue Zhao}

\author[1,\Letter]{Jianjun Xu}

\author[2]{Haoyang Cheng}

\email{wqcui@ustc.edu.cn}
\email{xjj1994@mail.ustc.edu.cn}

\runningtitle{Online Kernel Sliced Inverse Regression}
\runningauthor{Author-a \etal}

\address{International Institute of Finance, School of Management, University of Science and Technology of China, Hefei 230026, China}

\address{College of Electrical and Information Engineering, Quzhou University, Quzhou 324000, China}

\GraphicalAbstract{
}

\abstract{
  Online dimension reduction is a common method for high-dimensional streaming data processing. Online principal component analysis, online sliced inverse regression, online kernel principal component analysis and other methods have been studied in depth, but as far as we know, online supervised nonlinear dimension reduction methods have not been fully studied. 
  In this article, an online kernel sliced inverse regression method is proposed. 
  By introducing the approximate linear dependence condition and dictionary variable sets, we address the problem of increasing variable dimensions with the sample size in the online kernel sliced inverse regression method, and propose a reduced-order method for updating variables online.
  We then transform the problem into an online generalized eigen-decomposition problem, and use the stochastic optimization method to update the centered dimension reduction directions. Simulations and the real data analysis show that our method can achieve close performance to batch processing kernel sliced inverse regression. 
}


\begin{document}

\maketitle 

\section{Introduction}

With the arrival of big data and cloud computing, the processing of high-dimensional streaming data becomes a hot issue in machine learning. 
In practical data analysis, the high-dimensionality of data brings a series of problems, often referred to as the ``curse of dimensionality''. 
A common approach to dealing with high-dimensional data is first to reduce the dimension, and then perform statistical analysis on the lower dimensional data. 
Our goal is to reduce the dimension of the streaming data into a range suitable for further analysis, while minimizing the loss of information during the dimension reduction process. 
Consider the response variable $y \in \mathcal{R}$ and the predictor $\mathbf{x}=\left(x_{1}, \dots, x_{p}\right)^{\top} \in \mathcal{R}^{p}$. 
If there exists $d(d<p)$ dimensional statistic $S(\mathbf{x})$ satisfying the conditional independence assumption: $y \perp \! \! \! \perp \mathbf{x} \vert S(\mathbf{x})$, where $\perp \! \! \! \perp$ means that $y$ is statistically independent of $\mathbf{x}$, then we only need to obtain the summary statistic $S(\mathbf{x})$ to make statistical inference on $y$. 

When  $S(\mathbf{x})$ is linear with respect to $\mathbf{x}$, then $S(\mathbf{x}) = (\boldsymbol{\beta}_{1}^{\top} \mathbf{x}, \dots, \boldsymbol{\beta }_{d}^{\top} \mathbf{x})^{\top}$ where $\boldsymbol{\beta}_{1}, \dots, \boldsymbol{\beta}_{d}$
are referred to as effective dimension reduction~(e.d.r.) directions and the $d$-dimensional subspace $\mathrm{span}(\boldsymbol{\beta}_{1}, \dots, \boldsymbol{\beta}_{d})$ is defined as the e.d.r. subspace. 
Under  the linear design condition presented by Li~\cite{Li1991SIR}, the e.d.r. directions correspond to the eigenvectors of the matrix $\cov(\mathbf{x})^{-1} \cov(\mathbb{E}[\mathbf{x}|y])$. 
Sliced Inverse Regression~(SIR)~\cite{Li1991SIR}, Sliced Average Variance Estimation~(SAVE)~\cite{Cook2000SAVE}, Principal Hessian Directions~(PHD)~\cite{Li1992PHD} and other methods have been studied for the estimation of linear e.d.r subspaces. 
A common assumption in the analysis of high-dimensional data is that the intrinsic structure of the data is actually low-dimensional, e.g., the data is concentrated on some manifold. 
In the nonlinear case, the above linear methods do not characterize the inner structure well. 

These limits prompted Wu~\cite{WuHM} to propose a nonlinear extension of SIR called  Kernel Sliced Inverse Regression (KSIR), based on the kernel method~\cite{Ajzerman, Hastie} in statistical learning. The original data is mapped into a feature space
induced by the reproducing kernel.
By applying SIR to this feature space, we can obtain a set of linear directions in this feature space 
and this set of directions reflects the nonlinear structure in the original data space. 
The low-dimensional statistic $S(\mathbf{x})$ obtained by KSIR can be further analyzed for prediction or classification. 
Yeh et al.~\cite{Yeh} further enhanced the theoretical basis of KSIR by embedding the original data $\mathbf{x}$ into a feature space through an embedding map $\phi$, proving that this feature space is isometrically isomorphic to a reproducing kernel Hilbert space~(RKHS), and using the properties of RKHS to give a better theoretical interpretation of KSIR. 
Wu et al.~\cite{Wu} further extended the theory of KSIR by using the concept of the Mercer kernel and used two types of regularization schemes to mitigate the computational instability arising in KSIR. 

The need for immediate analysis and real-time feedback makes the above-mentioned batch processing dimension reduction methods ineffective when high-dimensional data arrives in a stream. 
To this need,  several online dimension reduction methods were proposed. 
Principal Component Analysis~(PCA)~\cite{Pearson1901pca} is the most common dimension reduction method, and some researchers  have proposed different online PCA methods~\cite{hall1998incremental,weng2003candid,arora2012stochastic}. 
In the field of sufficient dimension reduction, Zhang and Wu~\cite{zhang2019online} extended the incremental PCA to incremental SIR. 
Cai et al.~\cite{cai2020online} proposed two online SIR methods based on the perturbation and gradient descent methods. 
Xu et al.~\cite{cheng2021online} proposed an online sparse SIR method using the truncated gradient method. 
The research on online nonlinear dimension reduction methods mainly focuses on the online version of Kernel Principal Component Analysis~(KPCA)~\cite{scholkopf1997kernel}. 
Kimura et al.~\cite{kimura2005incremental} proposed the online KPCA method with the incremental update. 
Honeine~\cite{honeine2011online} proposed a reduced-order online KPCA model. 
Ghashami et al.~\cite{ghashami2016streaming} used a random feature mapping approach for online KPCA. 

In this article, we propose an Online Kernel Sliced Inverse Regression~(OKSIR) method, which is the online fashion of KSIR. 
The first difficulty of the OKSIR method is that the dimension of the kernel matrix increases as the amount of data increases, 
which leads to an explosion of computational effort. 
In this paper, we adopt the method proposed by Engel et al.~\cite{engel2004kernel} to construct a small subset of the training samples --- the dictionary during the streaming training process. 
A newly arrived sample will be added to the dictionary based on its distance from the previously stored dictionary samples in the feature space. 
For the samples that are not added to the dictionary, we represent these samples by linear combinations of dictionary samples and obtain the reduced-order estimation of the kernel matrix from these linear representations, so that we solve the problem of increasing dimensionality of the kernel matrix. 
Another difficulty is the computation of the slice matrix with respect to the reduced-order kernel matrix. 
We address this problem by treating the product of the slice matrix and the coefficient matrix as a whole to be updated. 
Based on the above reduced-order representation of the kernel matrix, we can transform the online kernel sliced inverse regression problem into an online updated generalized eigen-decomposition problem and solve it by stochastic optimization~\cite{chen2019online}. 

The paper is structured as follows. Section \ref{sec2} briefly reviews the kernel sliced inverse regression method and gives the approximate linear dependence~(ALD) condition. 
Section \ref{sec3} introduces our proposed online kernel sliced inverse regression method. 
Section \ref{sec4} presents the simulation experiments and real data analysis. 
Section \ref{sec5} finishes the article with a brief conclusion. 

\section{KSIR and ALD condition}

\label{sec2}

\subsection{Review of Kernel Sliced Inverse Regression}

We first briefly review the kernel sliced inverse regression method. 
Consider the predictor $\mathbf{x} \in \mathcal{X} \subseteq \mathcal{R}^{p}$, and the response variable $y \in \mathcal{Y} \subseteq \mathcal{R}$. 
Given a Mercer kernel function $k(\cdot, \cdot)$, one can construct an embedding $\phi$ that maps $\mathbf{x}$ to a feature Hilbert space $\mathcal{H}$, which is defined by the eigenvalues and eigenfunctions of the Mercer kernel~\cite{mercer1909functions}. 
The dimension of this Hilbert space may be infinite, and its inner product is $\langle\phi(\mathbf{x}), \phi(\mathbf{z})\rangle_{\mathcal{H}}=k(\mathbf{x}, \mathbf{z})$ for $\mathbf{x}, \mathbf{z}\in \mathcal{X}$. 
Hereafter, we substitute $\langle\cdot, \cdot\rangle$ for $\langle\cdot, \cdot\rangle_{\mathcal{H}}$ for notational simplicity. 

Consider the following regression model in the feature space~\cite{WuHM}:
\begin{equation}
\label{eq1}
y=f\left(\left\langle\boldsymbol{\beta}_{1}, \phi(\mathbf{x})\right\rangle,\left\langle\boldsymbol{\beta}_{2}, \phi(\mathbf{x})\right\rangle, \ldots,\left\langle\boldsymbol{\beta}_{d}, \phi(\mathbf{x})\right\rangle, \varepsilon\right), 
\end{equation}
where $\boldsymbol{\beta}_j \in \mathcal{H}$, and the distribution of $\varepsilon$ is independent of $\mathbf{x}$. 
Assume that the features satisfy the linear design condition: 
\begin{condition}
\label{cond1}
For any $f \in \mathcal{H}$, suppose there exists $\mathbf{b} \in \mathcal{R}^{d}$ satisfies $\mathbb{E}[\langle f, \phi(\mathbf{x})\rangle \vert S(\mathbf{x})] = \mathbf{b}^{\top} S(\mathbf{x})$ with $S(\mathbf{x}) = \left(\left\langle\boldsymbol{\beta}_{1}, \phi(\mathbf{x})\right\rangle,\left\langle\boldsymbol{\beta}_{2}, \phi(\mathbf{x})\right\rangle, \ldots,\left\langle\boldsymbol{\beta}_{d}, \phi(\mathbf{x})\right\rangle\right)^{\top}$. 
\end{condition}
Although $S(\mathbf{x})$ here is a linear statistic in the feature space, it corresponds to a nonlinear structure in $\mathcal{X}$. We define 
$\boldsymbol{\beta}_1,\cdots,\boldsymbol{\beta}_d$ the nonlinear e.d.r. directions and $\mathrm{span}(\boldsymbol{\beta}_1, \dots, \boldsymbol{\beta}_{d})$ the nonlinear e.d.r. subspace. 
Under model (\ref{eq1}) and Condition \ref{cond1}, the inverse regression curve 
$\mathbb{E}[\phi(\mathbf{x}) \vert y]-\mathbb{E}[\phi(\mathbf{x})]$ falls into the span of $(\boldsymbol{\Sigma} \boldsymbol{\beta}_{1}, \dots, \boldsymbol{\Sigma} \boldsymbol{\beta}_{d})$, where $\boldsymbol{\Sigma}=\cov[\phi(\mathbf{x})]$ is the covariance operator of $\phi(\mathbf{x})$. 
As a result, the sliced inverse regression method in the feature space requires only the solution of the following generalized eigen-decomposition problem 
\begin{equation}
\label{eq2}
\boldsymbol{\Gamma} \boldsymbol{\beta}=\lambda \boldsymbol{\Sigma} \boldsymbol{\beta},
\end{equation}
where $\boldsymbol{\Gamma}=\cov[\mathbb{E}[\phi(\mathbf{x}) \vert y]]$. 

Consider the data set $\left\{\left(\mathbf{x}_{i}, y_{i}\right), i \in 1,2, \dots, n\right\}$, and let $\mathbf{K}$ be the kernel matrix defined by the kernel function $k(\cdot, \cdot)$ with respect to $\mathbf{x}_1, \mathbf{x}_2, \dots, \mathbf{x}_n$, and its $(i, j)$-th element $\mathbf{K}_{i, j}=\left\langle\phi\left(\mathbf{x}_{i}\right), \phi\left(\mathbf{x}_{j}\right)\right\rangle=k\left(\mathbf{x}_{i}, \mathbf{x}_{j}\right), i, j=1,2, \dots, n$. 
$\mathbf{K}$ is also known as the Gram matrix in some other literature. 
Wu~\cite{WuHM} proved that (\ref{eq2}) is equivalent to the following generalized eigen-decomposition problem 
\begin{equation}
\label{eq3}
\mathbf{K J K c}=\lambda \mathbf{K}^2 \mathbf{c} , 
\end{equation}
where $\mathbf{J}$ is a $n \times n$ slice matrix, which is constructed by dividing the samples into $H$ slices according to the order statistics of $y$, the $h$-th slice contains $n_{h}$ samples. 
We partition the range of $y$ into $H$ slices, $I_1, \dots, I_H$, where $I_h = (q_{h-1}, q_{h}]$, and $q_h$'s are pre-specified cutting points satisfying $-\infty = q_0 < q_1 < \dots < q_H = \infty$. 
When the $i, j$ items are in the same $h$-th slice, $\mathbf{J}_{i, j}=1 / n_{h}$, otherwise $\mathbf{J}_{i, j}=0$. 

Take a look at model (\ref{eq1}), we can see that subsequent analysis does not require the specific form of $\phi(\mathbf{x})$, but only needs to estimate the $d$ summary statistics $v_{j}=\left \langle \boldsymbol{\beta}_{j}, \phi(\mathbf{x}) \right \rangle, j=1,2, \dots, d$. 
The following theorem establishes the connection between the two eigen-decomposition problems and gives one way to obtain the estimation of $v_1,\ldots,v_d$. 

\begin{theorem}
\label{thm1}
\cite{WuHM}~Given observations $\left\{\left(\mathbf{x}_{1}, y_{1}\right), \ldots,\left(\mathbf{x}_{n}, y_{n}\right)\right\}$, let $(\hat{\boldsymbol{\beta}}_{1}, \hat{\boldsymbol{\beta}}_{2}, \dots, \hat{\boldsymbol{\beta}}_{d})$ and $\left(\hat{\mathbf{c}}_{1}, \hat{\mathbf{c}}_{2}, \dots, \hat{\mathbf{c}}_{d}\right)$ denote the eigenvectors of (\ref{eq2}) and (\ref{eq3}). 
Then for any $\mathbf{x} \in \mathcal{X}$ and $j=1,2, \dots, d$, when the sample covariance $\widehat{\boldsymbol{\Sigma}} = \frac{1}{n} \sum_{i=1}^n \phi\left(\mathbf{x}_i\right) \otimes \phi\left(\mathbf{x}_i\right)$ is invertible, the following equation holds:
$$\quad \hat{v}_{j}=\left\langle\hat{\boldsymbol{\beta}}_{j}, \phi(\mathbf{x})\right\rangle=\hat{\mathbf{c}}_{j}^{\top} \mathbf{K}_{x} , $$
where $\mathbf{K}_{x}=\left(k\left(\mathbf{x}, \mathbf{x}_{1}\right), k\left(\mathbf{x}, \mathbf{x}_{2}\right), \dots, k\left(\mathbf{x}, \mathbf{x}_{n}\right)\right)^{\top}$. 
If $\widehat{\boldsymbol{\Sigma}}$ is not invertible, the conclusion still holds when only the image space of $\widehat{\boldsymbol{\Sigma}}$ is considered. 
\end{theorem}

Theorem~\ref{thm1} states that the statistics $v_1, v_2, \dots, v_d$ can be calculated as soon as the solution to problem \eqref{eq3} is obtained. 
Therefore, it suffices to study the online solution of problem\eqref{eq3}. 

\subsection{Approximate Linear Dependence Condition}

In this subsection, we focus on the approximate linear dependence~(ALD) condition. 
As Theorem~\ref{thm1} shows, we need to update $\mathbf{K}$ and $\mathbf{J}$ online, while the sizes of $\mathbf{K}$ and $\mathbf{J}$ will increase with the amount of data, which significantly increases the computational burden of batch processing methods. 
Given the data stream $\left\{\left(\mathbf{x}_{i}, y_{i}\right), i=1,2, \cdots, t\right\}$ up to the $t$-th time step, we can get $\hat{v}_{j} = \sum_{i=1}^{t} \hat{c}_{j, i} k(\mathbf{x}, \mathbf{x}_i) = \sum_{i=1}^{t} \hat{c}_{j, i} \left\langle\phi\left(\mathbf{x}_{i}\right), \phi(\mathbf{x})\right\rangle$, where $\hat{c}_{j,i}$ is the $i$-th element of $\hat{\mathbf{c}}_j$. 
If we can express $\phi\left(\mathbf{x}_{t}\right)$ as $\phi\left(\mathbf{x}_{t}\right)=\sum_{i=1}^{t-1} a_{i} \phi\left(\mathbf{x}_{i}\right)$, then the coefficient $\hat{c}_{j,t}$ in the expression of $\hat{v}_{j}$ can be reduced to zero. 
When $\operatorname{dim}(\mathcal{H})$ is finite, the above expression holds. 
We can use the first $\operatorname{dim}(\mathcal{H})$ linearly independent vectors of the data stream as a basis, and this basis can linearly represent all subsequent samples. 
When $\operatorname{dim}(\mathcal{H})=\infty$, Engel et al.~\cite{engel2004kernel} define the ALD condition to measure the linear dependence between $\phi\left(\mathbf{x}_{t}\right)$ and the previous samples. 
Suppose that at $t$-th time step, we have collected a dictionary consisting of a subset of $m_{t-1}$ training samples $\mathcal{D}_{t-1}=\left\{\widetilde{\mathbf{x}}_{j}\right\}_{j=1}^{m_{t-1}}$, where $\left\{\phi\left(\widetilde {\mathbf{x}}_{j}\right)\right\}_{j=1}^{m_{t-1}}$ is a set of linearly independent feature vectors, called dictionary vectors. 
For a newly come sample $\mathbf{x}_t$, 
to determine whether $\mathbf{x}_{t}$ should be added to the dictionary, we test whether $\phi\left(\mathbf{x}_{t}\right)$ is approximately linearly dependent on the dictionary vectors. 
If not, we add it to the dictionary. 

The square approximation error of $\phi(\mathbf{x}_t)$ by a linear combination of $\mathcal{D}_{t-1}$ is given by 
\begin{equation*}
\epsilon_t:=\min_{\mathbf{a}}\left\|\sum_{j=1}^{m_{t-1}} a_{j} \phi\left(\widetilde{\mathbf{x}}_{j}\right)-\phi\left(\mathbf{x}_{t}\right)\right\|^{2}.
\end{equation*}
By expanding this norm, we have the matrix form of the approximation error 
\begin{equation*}
\begin{gathered}
\epsilon_t:=\min_{\mathbf{a}} \mathbf{a}^{\top} \widetilde{\mathbf{K}}_{t-1} \mathbf{a} - 2 \mathbf{a}^{\top} \widetilde{\mathbf{k}}_{t-1}(\mathbf{x}_t) + k_{t t}, \\ 
\text {with } 
{\left[\widetilde{\mathbf{K}}_{t-1}\right]_{i, j}=k\left(\widetilde{\mathbf{x}}_{i}, \widetilde{\mathbf{x}}_{j}\right),\left(\widetilde{\mathbf{k}}_{t-1}\left(\mathbf{x}_{t}\right)\right)_{i}=
k\left(\widetilde{\mathbf{x}}_{i}, \mathbf{x}_{t}\right),} \\
k_{t t}=k\left(\mathbf{x}_{t}, \mathbf{x}_{t}\right), i, j=1,2, \ldots, m_{t-1} . 
\end{gathered}
\end{equation*}
By taking the derivative of the above cost function with respect to $\mathbf{a}$ and setting it to zero, we get the optimal solution
\begin{equation}
\label{eq4}
\begin{gathered}
\widetilde{\mathbf{a}}_{t}=\widetilde{\mathbf{K}}_{t-1}^{-1} \widetilde{\mathbf{k}}_{t-1}\left(\mathbf{x}_{t}\right), \\
\epsilon_t=k_{t t}-\widetilde{\mathbf{k}}_{t-1}\left(\mathbf{x}_{t}\right)^{\top} \widetilde{\mathbf{a}}_{t}. 
\end{gathered}
\end{equation}

Now that we get the approximation error $\epsilon_t$, given a threshold parameter $\nu$, we have the following ALD condition. 
\begin{definition}
\label{def1}
Upon the arrival of $\mathbf{x}_t$, we add $\mathbf{x}_t$ into the dictionary if $$\epsilon_t > \nu . $$
Otherwise,  the dictionary remains unchanged. 
\end{definition}

According to Definition \ref{def1}, $\phi(\mathbf{x}_t)$ can be expressed as $$\phi(\mathbf{x}_t) = \sum_{j=1}^{m_t} a_{t, j} \phi(\widetilde{\mathbf{x}}_j) + \phi_{t}^{res},  $$where $\phi_{t}^{res}$ denotes the residual error vector. 
If $\epsilon_t \leq \nu$, then $\mathbf{x}_t$ satisfies the ALD condition on $\mathcal{D}_{t-1}$, we will not augment $\mathbf{x}_t$ into the dictionary. 
In this case, $\mathbf{a}_t = \widetilde{\mathbf{a}}_t$, and $\|\phi_{t}^{res}\|^2 \leq \nu$. 
Otherwise, $\mathbf{a}_t = (0, \dots, 0, 1)^{\top}$ and $\phi_{t}^{res} = 0$. 
By choosing $\nu$ sufficiently small, we can make sure the approximation error of $\phi(\mathbf{x}_t) \approx \sum_{j=1}^{m_t} a_{t, j} \phi(\widetilde{\mathbf{x}}_j)$ is correspondingly small. 
By transforming this approximation into matrix notation, we have $\mathbf{K}_t \approx \mathbf{A}_t \widetilde{\mathbf{K}}_t \mathbf{A}_t^{\top}$, where $\left[\mathbf{K}_{t}\right]_{i, j}=k\left(\mathbf{x}_{i}, \mathbf{x}_{j}\right), i, j=1, \ldots, t,\left[\mathbf{A}_{t}\right]_{i, j}=a_{i, j}$, $i=1, \ldots, t$,  $j=1, \ldots, m_{t}$. 
To solve the generalized eigen-decomposition problem \eqref{eq3} in an online fashion, we replace $\mathbf{K}$ in equation \eqref{eq3} with $\mathbf{K}_t = \mathbf{A}_t \widetilde{\mathbf{K}}_t \mathbf{A}_t^{\top}$ which yields
\begin{equation*}
\mathbf{A}_{t} \widetilde{\mathbf{K}}_{t} \mathbf{A}_{t}^{\top} \mathbf{J}_{t} \mathbf{A}_{t} \widetilde{\mathbf{K}}_{t} \mathbf{A}_{t}^{\top} \mathbf{c}=\lambda \mathbf{A}_{t} \widetilde{\mathbf{K}}_{t} \mathbf{A}_{t}^{\top} \mathbf{A}_{t} \widetilde{\mathbf{K}}_{t} \mathbf{A}_{t}^{\top} \mathbf{c} . \\
\end{equation*}
Let $\boldsymbol{\alpha}=\mathbf{A}_{t}^{\top} \mathbf{c}$, $\mathbf{Q}_t = \mathbf{A}_{t}^{\top} \mathbf{J}_{t} \mathbf{A}_{t}$ and eliminate $\mathbf{A}_t$ from both sides, we have 
\begin{equation}
\label{eq5}
\widetilde{\mathbf{K}}_{t} \mathbf{Q}_{t} \widetilde{\mathbf{K}}_{t} \boldsymbol{\alpha}=\lambda \widetilde{\mathbf{K}}_{t} \mathbf{A}_{t}^{\top} \mathbf{A}_{t} \widetilde{\mathbf{K}}_{t} \boldsymbol{\alpha}.
\end{equation}
Let $\left(\boldsymbol{\alpha}_1, \boldsymbol{\alpha}_2, \dots, \boldsymbol{\alpha}_d\right)$ denote the eigenvectors of (\ref{eq5}), $v_j$ can be estimated as 
\begin{equation*}
\begin{aligned}
\hat{v}_{j} &=\hat{\mathbf{c}}_{j}^{\top}\left(\left\langle\phi\left(\mathbf{x}_{1}\right), \phi(\mathbf{x})\right\rangle, \ldots,\left\langle\phi\left(\mathbf{x}_{t}\right), \phi(\mathbf{x})\right\rangle\right)^{\top} \\
&=\hat{\mathbf{c}}_{j}^{\top}\left(\sum_{j=1}^{m_{t}} a_{1, j}\left\langle\phi\left(\widetilde{\mathbf{x}}_{j}\right), \phi(\mathbf{x})\right\rangle, \ldots, \sum_{j=1}^{m_{t}} a_{t, j}\left\langle\phi\left(\widetilde{\mathbf{x}}_{j}\right), \phi(\mathbf{x})\right\rangle\right)^{\top} \\
&=\hat{\mathbf{c}}_{j}^{\top} \mathbf{A}_{t} \widetilde{\mathbf{k}}_{t}(\mathbf{x}) \\
&=\hat{\boldsymbol{\alpha}}_{j}^{\top} \widetilde{\mathbf{k}}_{t}(\mathbf{x}) . 
\end{aligned}
\end{equation*}

\section{Online Kernel Sliced Inverse Regression Method}

\label{sec3}

\subsection{Online Update for the Variables}

During the online update, we will encounter two cases depending on the comparison between $\epsilon_t$ and $\nu$. 
\begin{itemize}[leftmargin=*]
    \item \textbf{Case 1} $\epsilon_t \leq \nu$. In this case, the dictionary remains unchanged, and $\mathcal{D}_{t}=\mathcal{D}_{t-1}, m_{t}=m_{t-1}, \widetilde{\mathbf{K}}_{t}=\widetilde{\mathbf{K}}_{t-1}$. 
    The coefficients $\mathbf{a}_{t}=\widetilde{\mathbf{a}}_{t}$, $\mathbf{A}_{t}=\left[\mathbf{A}_{t-1}^{\top}, \mathbf{a}_{t}\right]^{\top}$, $\mathbf{A}_{t}^{\top} \mathbf{A}_{t}=\mathbf{A}_{t-1}^{\top} \mathbf{A}_{t-1}+\mathbf{a}_{t} \mathbf{a}_{t}^{\top}$. 
    As for $\mathbf{Q}_t$, we have 
    $$
\begin{aligned}
\mathbf{Q}_{t} &=\mathbf{A}_{t}^{\top} \mathbf{J}_{t} \mathbf{A}_{t}=\sum_{h=1}^{H} \frac{1}{n_{h, t}} \mathbf{A}_{t}^{\top} \boldsymbol{\Delta}_{h, t} \mathbf{\Delta}_{h, t}^{\top} \mathbf{A}_{t} \\
&=\sum_{h=1}^{H} \frac{1}{n_{h, t}}\left[\mathbf{A}_{t-1}^{\top}, \mathbf{a}_{t} \right]\left[\boldsymbol{\Delta}_{h, t-1}^{\top},    \delta_{h}\left(y_{t}\right)\right]^{\top} {\left[\boldsymbol{\Delta}_{h, t-1}^{\top},    \delta_{h}\left(y_{t}\right)\right]\left[\mathbf{A}_{t-1}^{\top}, \mathbf{a}_{t} \right]^{\top} } \\
&=\sum_{h=1}^{H} \frac{1}{n_{h, t}}\left(\mathbf{A}_{t-1}^{\top}    \boldsymbol{\Delta}_{h, t-1}+\delta_{h}\left(y_{t}\right) \mathbf{a}_{t}\right)  \left(\boldsymbol{\Delta}_{h, t-1}^{\top} \mathbf{A}_{t-1}+\delta_{h}\left(y_{t}\right)    \mathbf{a}_{t}^{\top}\right) \\
&=\sum_{h=1}^{H} \frac{1}{n_{h, t}}\left(\mathbf{A}_{t-1}^{\top}    \boldsymbol{\Delta}_{h, t-1} \mathbf{\Delta}_{h, t-1}^{\top} \mathbf{A}_{t-1}+\right.\\
&\left.\delta_{h}\left(y_{t}\right)\left(\mathbf{a}_{t} \mathbf{\Delta}_{h, t-1}^{\top} \mathbf{A}_{t-1}+\mathbf{A}_{t-1}^{\top} \boldsymbol{\Delta}_{h, t-1}    \mathbf{a}_{t}^{\top}+\mathbf{a}_{t} \mathbf{a}_{t}^{\top}\right)\right) \\
&=\sum_{h=1}^{H} \frac{1}{n_{h, t}}\left(\mathbf{M}_{h, t-1}+\delta_{h}\left(y_{t}\right)\left(\mathbf{a}_{t} \mathbf{m}_{h,    t-1}^{\top}+\mathbf{m}_{h, t-1} \mathbf{a}_{t}^{\top}+\mathbf{a}_{t}    \mathbf{a}_{t}^{\top}\right)\right) \\
&=\sum_{h=1}^{H} \frac{1}{n_{h, t}} \mathbf{M}_{h, t},
\end{aligned}
$$
where $\boldsymbol{\Delta}_{h, t}=\left[\delta_{h}\left(y_{1}\right), \ldots, \delta_{h}\left(y_{t}\right)\right]^{\top}$, and $\delta_{h}\left(y_{t}\right)$ is the indicator variable indicating whether $y_{t}$ belongs to the $h$-th slice or not. 
When $y_{t}$ belongs to the $h$-th slice, $\delta_{h}\left(y_{t}\right) = 1$, otherwise, it equals to zero. 
The update formula for intermediate variables is shown below
\begin{equation}
\label{eq6}
\begin{aligned}
&n_{h, t}=n_{h, t-1}+\delta_{h}\left(y_{t}\right)  , \\
&\boldsymbol{\Delta}_{h, t}=\left[\boldsymbol{\Delta}_{h, t-1}, \delta_{h}\left(y_{t}\right)\right]^{\top} , \\
&\mathbf{M}_{h, t}=\mathbf{M}_{h, t-1}+\delta_{h}\left(y_{t}\right)\left(\mathbf{a}_{t} \mathbf{m}_{h, t-1}^{\top}+\mathbf{m}_{h, t-1} \mathbf{a}_{t}^{\top}+\mathbf{a}_{t} \mathbf{a}_{t}^{\top}\right) , \\
&\mathbf{m}_{h, t}=\mathbf{m}_{h, t-1}+\delta_{h}\left(y_{t}\right) \mathbf{a}_{t} . 
\end{aligned}
\end{equation}
    \item \textbf{Case 2} $\epsilon_t > \nu$. 
    $\mathbf{x}_{t}$ is added to the dictionary, which means $\mathcal{D}_{t}=\mathcal{D}_{t-1} \cup\left\{\mathbf{x}_{t}\right\}$, $m_{t}=m_{t-1}+1$. For $\widetilde{\mathbf{K}}_{t}$, we have 
\begin{equation}
\label{eq7}
\begin{aligned}
\widetilde{\mathbf{K}}_{t} &=\left[\begin{array}{ll}
\widetilde{\mathbf{K}}_{t-1} & \widetilde{\mathbf{k}}_{t-1}\left(\mathbf{x}_{t}\right) \\
\widetilde{\mathbf{k}}_{t-1}\left(\mathbf{x}_{t}\right)^{\top} & k_{t t}
\end{array}\right] \Rightarrow \\
\widetilde{\mathbf{K}}_{t}^{-1} &=\frac{1}{\epsilon_t}\left[\begin{array}{ll}
\epsilon_t \widetilde{\mathbf{K}}_{t-1}^{-1}+\widetilde{\mathbf{a}}_{t} \widetilde{\mathbf{a}}_{t}^{\top} & -\widetilde{\mathbf{a}}_{t} \\
-\widetilde{\mathbf{a}}_{t}^{\top} & 1
\end{array}\right] , 
\end{aligned}
\end{equation}
where $\widetilde{\mathbf{a}}_{t}=\widetilde{\mathbf{K}}_{t-1}^{-1} \widetilde{\mathbf{k}}_{t-1}\left(\mathbf{x}_{t}\right)$ is the solution in \eqref{eq4}. 
Since $\mathbf{x}_{t}$ is a new dictionary sample, $\mathbf{a}_{t}=(0, \ldots, 0,1)^{\top}$, 
we have 
\begin{equation}
\label{eq8}
\begin{aligned}
\mathbf{A}_{t} & = \left[\begin{array}{ll}
\mathbf{A}_{t-1} & \mathbf{0} \\
\mathbf{0}^{\top} & 1
\end{array}\right], \\
\mathbf{A}_{t}^{\top} \mathbf{A}_{t} & = \left[\begin{array}{ll}
\mathbf{A}_{t-1}^{\top} \mathbf{A}_{t-1} & \mathbf{0} \\
\mathbf{0}^{\top} & 1
\end{array}\right] . 
\end{aligned}
\end{equation}
As for $\mathbf{Q}_t$, we have 
$$
\begin{aligned}
&\mathbf{Q}_{t}=\mathbf{A}_{t}^{\top} \mathbf{J}_{t} \mathbf{A}_{t}=\sum_{h=1}^{H} \frac{1}{n_{h, t}} \mathbf{A}_{t}^{\top} \boldsymbol{\Delta}_{h, t} \boldsymbol{\Delta}_{h, t}^{\top} \mathbf{A}_{t}\\
&=\sum_{h=1}^{H} \frac{1}{n_{h, t}}\left[\begin{array}{ll}
\mathbf{A}_{t-1} & \mathbf{0} \\
\mathbf{0}^{\top} & 1
\end{array}\right]^{\top}\left[\begin{array}{l}
\boldsymbol{\Delta}_{h, t-1} \\
\delta_{h}\left(y_{t}\right)
\end{array}\right] 
\left[\boldsymbol{\Delta}_{h, t-1}^{\top}, \delta_{h}\left(y_{t}\right)\right]\left[\begin{array}{ll}
\mathbf{A}_{t-1} & \mathbf{0} \\
\mathbf{0}^{\top} & 1
\end{array}\right]\\
&=\sum_{h=1}^{H} \frac{1}{n_{h, t}}
\left[\begin{array}{ll}
\mathbf{A}_{t-1}^{\top} \boldsymbol{\Delta}_{h, t-1} \boldsymbol{\Delta}_{h, t-1}^{\top} \mathbf{A}_{t-1} & \mathbf{A}_{t-1}^{\top} \boldsymbol{\Delta}_{h, t-1} \delta_{h}\left(y_{t}\right) \\
\delta_{h}\left(y_{t}\right) \boldsymbol{\Delta}_{h, t-1}^{\top} \mathbf{A}_{t-1} & \delta_{h}\left(y_{t}\right)
\end{array}\right]\\
&=\sum_{h=1}^{H} \frac{1}{n_{h, t}}\left[\begin{array}{ll}
\mathbf{M}_{h, t-1} & \mathbf{m}_{h, t-1} \delta_{h}\left(y_{t}\right) \\
\delta_{h}\left(y_{t}\right) \mathbf{m}_{h, t-1}^{\top} & \delta_{h}\left(y_{t}\right)
\end{array}\right]\\
&=\sum_{h=1}^{H} \frac{1}{n_{h, t}} \mathbf{M}_{h, t} \text {. }
\end{aligned}
$$
The update formula for intermediate variables is shown below
\begin{equation}
\label{eq9}
\begin{aligned}
&n_{h, t}=n_{h, t-1}+\delta_{h}\left(y_{t}\right)  , \\
&\boldsymbol{\Delta}_{h, t}=\left[\boldsymbol{\Delta}_{h, t-1}, \delta_{h}\left(y_{t}\right)\right]^{\top} , \\
&\mathbf{M}_{h, t}=\left[\begin{array}{ll}
\mathbf{M}_{h, t-1} & \mathbf{m}_{h, t-1} \delta_{h}\left(y_{t}\right) \\
\mathbf{m}_{h, t-1}^{\top} \delta_{h}\left(y_{t}\right) & \delta_{h}\left(y_{t}\right)
\end{array}\right] ,  \\
&\mathbf{m}_{h, t}=\left[\mathbf{m}_{h, t-1}^{\top}, \delta_{h}\left(y_{t}\right)\right]^{\top} . 
\end{aligned}
\end{equation}
\end{itemize}

\subsection{Online Kernel Sliced Inverse Regression Algorithm}

\begin{algorithm}[ht]
\caption{Online Kernel Sliced Inverse Regression Algorithm}
\label{alg1}
\begin{algorithmic}[1]
\Require the threshold parameter $\nu$
\Ensure the eigenvectors $\boldsymbol{\Phi}_t$
\State Initialize $\mathcal{D}_1 = \left\{\mathbf{x}_1\right\}$, $\mathbf{a}_1 = [1]$,  $\widetilde{\mathbf{K}}_1=\left[k_{11}\right], \widetilde{\mathbf{K}}_1^{-1}=\left[1 / k_{11}\right]$; $n_{h, 1} = \delta_h(y_1)$; $\boldsymbol{\Delta}_{h, 1} = \mathbf{M}_{h,1} = \mathbf{m}_{h,1} = [\delta_h(y_1)]$;  $\boldsymbol{\Phi}_1 \sim N_d(\mathbf{0}, 0.001\mathbf{I})$
\For{$t = 2, 3, \dots$}
    \State Get new sample $(\mathbf{x}_t, y_t)$; 
    \State Compute $\widetilde{\mathbf{k}}_{t-1}(\mathbf{x}_t)$; 
    \State Compute $\widetilde{\mathbf{a}}_t$ and $\epsilon_t$ (\ref{eq4}); 
    \If{$\epsilon_t \leq \nu$} \Comment{dictionary unchanged}
        \State $\mathcal{D}_{t}=\mathcal{D}_{t-1}$; 
        \State $\widetilde{\mathbf{K}}_{t}=\widetilde{\mathbf{K}}_{t-1}$, $\widetilde{\mathbf{K}}_{t}^{-1}=\widetilde{\mathbf{K}}_{t-1}^{-1}$; 
        \State $\mathbf{a}_{t}=\widetilde{\mathbf{a}}_{t}$, $\mathbf{A}_{t}^{\top} \mathbf{A}_{t}=\mathbf{A}_{t-1}^{\top} \mathbf{A}_{t-1}+\mathbf{a}_{t} \mathbf{a}_{t}^{\top}$; 
        \State Compute $n_{h,t}$, $\boldsymbol{\Delta}_{h,t}$, $\mathbf{m}_{h,t}$, $\mathbf{M}_{h,t}$ (\ref{eq6}); 
        \State $\mathbf{Q}_t = \sum_{h=1}^{H} \frac{1}{n_{h, t}} \mathbf{M}_{h, t}$; 
        \State Compute $\boldsymbol{\Phi}_t$ (\ref{eq10}); 
    \ElsIf{$\epsilon_t > \nu$} \Comment{add $\mathbf{x}_t$ to dictionary}
        \State $\mathcal{D}_{t}=\mathcal{D}_{t-1} \cup\left\{\mathbf{x}_{t}\right\}$; 
        \State Compute $\widetilde{\mathbf{K}}_t$ and $\widetilde{\mathbf{K}}_t^{-1}$ (\ref{eq7}); 
        \State Compute $\mathbf{A}_t^{\top} \mathbf{A}_t$ (\ref{eq8}); 
        \State Compute $n_{h,t}$, $\boldsymbol{\Delta}_{h,t}$, $\mathbf{m}_{h,t}$, $\mathbf{M}_{h,t}$ (\ref{eq9}); 
        \State $\mathbf{Q}_t = \sum_{h=1}^{H} \frac{1}{n_{h, t}} \mathbf{M}_{h, t}$; 
        \State Compute $\boldsymbol{\Phi}_t$ (\ref{eq11}); 
    \EndIf
\EndFor
\end{algorithmic}
\end{algorithm}

Now that we have the update formulas for $\widetilde{\mathbf{K}}_{t}$, $\mathbf{Q}_{t}$ and $\mathbf{A}_{t}^{\top} \mathbf{A}_{t}$ in \eqref{eq5}, 
we only need to update the solution of the generalized eigen-decomposition problem \eqref{eq5} online. 
Chen et al.~\cite{chen2019online} proposed a stochastic optimization algorithm for generalized eigen-decomposition problems, which does not require dual variables and does not involve the operation of matrix inversion. 
Let $\boldsymbol{\Phi}=\left(\boldsymbol{\alpha}_{1}, \boldsymbol{\alpha}_{2}, \dots, \boldsymbol{\alpha}_{d}\right)$ be the first $d$ eigenvectors of \eqref{eq5}. 

In \textbf{Case 1}, the ALD condition is satisfied. 
The update formula of $\boldsymbol{\Phi}$ is as follows: 
\begin{equation}
\label{eq10}
\boldsymbol{\Phi}_{t}=\boldsymbol{\Phi}_{t-1}-\eta_{t}\left(\widetilde{\mathbf{K}}_{t} \mathbf{Q}_{t} \widetilde{\mathbf{K}}_{t} \boldsymbol{\Phi}_{t-1} \boldsymbol{\Phi}_{t-1}^{\top}-\mathbf{I}_{m_{t}}\right) \widetilde{\mathbf{K}}_{t} \mathbf{A}_{t}^{\top} \mathbf{A}_{t} \widetilde{\mathbf{K}}_{t} \boldsymbol{\Phi}_{t-1} . 
\end{equation}

In \textbf{Case 2}, the ALD condition is not satisfied, and we first fill the dimension with $0$, 
\begin{equation}
\label{eq11}
\begin{aligned}
\boldsymbol{\Phi}_{t-} &=\left[\boldsymbol{\Phi}_{t-1}^{\top}, \mathbf{0}_{d}\right]^{\top} ,  \\
\boldsymbol{\Phi}_{t} &=\boldsymbol{\Phi}_{t^{-}}-\eta_{t}\left(\widetilde{\mathbf{K}}_{t} \mathbf{Q}_{t} \widetilde{\mathbf{K}}_{t} \boldsymbol{\Phi}_{t-} \boldsymbol{\Phi}_{t-}^{\top}-\mathbf{I}_{m_{t}}\right) \widetilde{\mathbf{K}}_{t} \mathbf{A}_{t}^{\top} \mathbf{A}_ {t} \widetilde{\mathbf{K}}_{t} \boldsymbol{\Phi}_{t-} . 
\end{aligned}
\end{equation}
$\eta_{t}$ is the learning rate, and we take $\eta_{t}=1 / t$ in  numerical studies. 
In practice, we can set the learning rate to $1/t$, then fix the learning rate to some $\eta$ after several steps. 
The algorithm in pseudo-code form is described in Algorithm \ref{alg1}. 

The time and space complexity of Algorithm \ref{alg1} at each step is also a concern for online learning. 
We first give a lemma that the cardinality of the dictionary is finite under certain conditions. 
\begin{lemma}
\label{lem1}
\cite{engel2004kernel}
Assume that the kernel function $k$ is a continuous Mercer kernel and the input space $\mathcal{X}$ is a compact subset of the Banach space. 
For any input sequence $\left\{\mathbf{x}_{t}\right\}_{t=1}^{\infty}$ and a threshold $\nu$ greater than 0, the cardinality of the dictionary variables is finite. 
\end{lemma}
Therefore, the number of dictionary samples $m_{t-1}$ is finite under certain conditions. 
Suppose that the upper bound of the number of dictionary samples is $m$. 
The time consumption of each step of our algorithm consists of two main parts. 
First is the variable update process, in which the most time-consuming part is the update of $\mathbf{Q}_{t}$ with the time complexity of $\mathcal{O}\left(m^{2} H\right)$. 
Second is the online update of the generalized eigen-decomposition problem, where the time complexity 
of computing the update of $\boldsymbol{\Phi}$ is $\mathcal{O}\left({m}^{2} d\right)$. 
In summary, the time complexity of our method at each step is $\mathcal{O}\left(m^{2}(d+H)\right)$. 

The space consumption of the algorithm is mainly for storing dictionary samples and  intermediate variables. 
The space consumption for storing dictionary samples is $\mathcal{O}(m p)$ and that for storing intermediate variables is $\mathcal{O}\left({m}^{2}\right)$. The total space complexity is $\mathcal{O} (m(m+p))$. 

\subsection{Properties of the OKSIR Method}

In this section, we will discuss some theoretical properties of our method. 
We start with some definitions and notations. 
Recall that $\mathbf{x}$ is a random vector in $\mathcal{X}$ and there exists an embedding $\phi: \mathcal{X} \rightarrow \mathcal{H}$ that maps $\mathbf{x}$ to a feature Hilbert space $\mathcal{H}$. 
Let $u$ be a random variable in $\mathcal{H}$ with $\mathbb{E} \|u\| < \infty$, where $\|\cdot\|$ is the norm in $\mathcal{H}$ introduced by its inner product $\left\langle \cdot,\cdot \right \rangle$. 
$\mathbb{E}\|u\|$ denotes the expectation of $u$, satisfying $\left \langle \mathbb{E}[u], v \right \rangle = \mathbb{E} [\left \langle u, v \right \rangle]$, $\forall v \in \mathcal{H}$. 
If $\mathbb{E}\|u\|^2 < \infty$, then the covariance of $u$ is defined to be $\cov(u) = \mathbb{E}[(u - \mathbb{E}[u]) \otimes (u - \mathbb{E}[u])]$. 
$\otimes$ denotes the tensor product in $\mathcal{H}$, and we have $(u \otimes v) w = \langle v, w \rangle u$, for all $w \in \mathcal{H}$. 

Let $\mathcal{P}$ be the measure for the random vector $\mathbf{x}$. 
We need the following technical assumption. 
\begin{assumption}
\label{assum1}
$\forall \mathbf{x} \in \mathcal{X}$, $k(\mathbf{x}, \cdot)$ is $\mathcal{P}$-measurable. 
There exists $M > 0$ such that $\mathbf{x} \in \mathcal{X}$, $k(\mathbf{x}, \mathbf{x}) \leq M$ (a.s.) with respect to $\mathcal{P}$. 
\end{assumption}
Assumption \ref{assum1} states that $\mathbb{E}\|\phi(\mathbf{x})\|^2 = \mathbb{E} [k(\mathbf{x}, \mathbf{x})]$ is bounded (a.s.). 
Thus $\phi(\mathbf{x})$ has a well-defined mean and covariance operator. 
Without loss of generality, we assume $\mathbb{E}[\phi(\mathbf{x})] = 0$ in this section. 
The covariance operator $\boldsymbol{\Sigma} = \cov[\phi(\mathbf{x})] = \mathbb{E}[\phi(\mathbf{x}) \otimes \phi(\mathbf{x})]$ is compact. 

Recall that $\boldsymbol{\Gamma}=\cov[\mathbb{E}[\phi(\mathbf{x}) \vert y]]$ is the covariance operator of the conditional expectation of $\phi(\mathbf{x})$ given  $y$. 
Define operator $\mathbf{T} = \boldsymbol{\Sigma}^{-1} \boldsymbol{\Gamma}$. 
The following proposition~\cite{Wu} describes the spectra of $\boldsymbol{\Gamma}$ and $\mathbf{T}$. 

\begin{proposition}
\label{prop1}
Under Assumption \ref{assum1} and Condition \ref{cond1}, we have, 
\begin{enumerate}
    \item the operator $\boldsymbol{\Gamma}$ has finite rank $d_{\Gamma} \leq d$. 
    Consequently, $\boldsymbol{\Gamma}$ is compact and has the following spectral decomposition 
    $$\boldsymbol{\Gamma}=\sum_{i=1}^{d_{\Gamma}} \tau_i \varphi_i \otimes \varphi_i , $$
    where $\tau_i$ and $\varphi_i$ are the eigenvalues and eigenvectors. 
    The eigenvectors $\varphi_i \in \mathbf{R}(\boldsymbol{\Sigma})$, $\forall i \in 1, 2, \dots, d_{\Gamma}$, where $\mathbf{R}(\boldsymbol{\Sigma})$ denotes the range of $\boldsymbol{\Sigma}$; 
    \item the generalized eigen-decomposition problem \eqref{eq2} is equivalent to the eigen-decomposition of $\mathbf{T}$, which takes the following form $$\mathbf{T} = \sum_{i=1}^{d_{\Gamma}} \tau_i \varphi_i \otimes \boldsymbol{\Sigma}^{-1} (\varphi_i) . $$
\end{enumerate}
\end{proposition}

Our goal is to establish an upper error bound for the estimation $\boldsymbol{\Phi}_t$. 
Without loss of generality, we assume that $\sum_{i=1}^{t} \phi(\mathbf{x}_i) = 0$, otherwise we can subtract $\bar{\phi}_t = \frac{1}{t} \sum_{i=1}^{t} \phi(\mathbf{x}_i)$ from $\phi(\mathbf{x}_i)$. 
The offline sample covariance at time t is estimated by 
\begin{equation*}
\widehat{\boldsymbol{\Sigma}}_t=\frac{1}{t} \sum_{i=1}^t \phi\left(\mathbf{x}_i\right) \otimes \phi\left(\mathbf{x}_i\right) . 
\end{equation*}
The sample covariance of $\mathbb{E}[\phi(\mathbf{x}) \vert y]$ can be estimated through slicing: 
\begin{equation*}
\widehat{\boldsymbol{\Gamma}}_t = \sum_{h=1}^{H} \frac{n_{h,t}}{t} \psi_{h,t} \otimes \psi_{h,t} , 
\end{equation*}
where $\psi_{h,t} = \frac{1}{n_{h,t}} \sum_{y_i \in I_h} \phi(\mathbf{x}_i)$ is the mean vector of each slice. 

In our OKSIR algorithm, the two covariance operators are replaced by their online versions. 
Recall that $\phi(\mathbf{x}_t)$ can be expressed as $\phi(\mathbf{x}_t) = \widetilde{\phi}(\mathbf{x}_t) + \phi_{t}^{res}$, where $\widetilde{\phi}(\mathbf{x}_t) = \sum_{j=1}^{m_t} a_{t, j} \phi(\widetilde{\mathbf{x}}_j)$. 
The online sample covariance has the following form 
\begin{equation*}
\widetilde{\boldsymbol{\Sigma}}_t=\frac{1}{t} \sum_{i=1}^t \widetilde{\phi}\left(\mathbf{x}_i\right) \otimes \widetilde{\phi}\left(\mathbf{x}_i\right) . 
\end{equation*}
Similarly, the online estimation of $\boldsymbol{\Gamma}$ can be written as 
\begin{equation*}
\widetilde{\boldsymbol{\Gamma}}_t = \sum_{h=1}^{H} \frac{n_{h,t}}{t} \widetilde{\psi}_{h,t} \otimes \widetilde{\psi}_{h,t} , 
\end{equation*}
where $\widetilde{\psi}_{h,t} = \frac{1}{n_{h,t}} \sum_{y_i \in I_h} \widetilde{\phi}(\mathbf{x}_i)$. 
 
\begin{lemma}
\label{lem2}
Under Assumption \ref{assum1} and Condition \ref{cond1}, $\forall \epsilon > 0$, there exists $T > 0$, for some constants $\delta_1$, $\delta_2$, $C_1$, $C_2$, $M_0 > 0$, for $t > T$, we have 
\begin{equation*}
\begin{aligned}
\mathrm{Pr} \left(\left\|\widetilde{\boldsymbol{\Sigma}}_t - \boldsymbol{\Sigma}\right\| > \frac{\delta_1}{\sqrt{t}} + \frac{C_1}{t} + M_0 \sqrt{\nu} + \nu \right) & < \epsilon ,  \\
\mathrm{Pr} \left(\left\|\widetilde{\boldsymbol{\Gamma}}_t - \boldsymbol{\Gamma}\right\| > \frac{\delta_2}{\sqrt{t}} + \frac{C_2}{t} + M_0 \sqrt{\nu} + \nu \right) & < \epsilon .  \\
\end{aligned}
\end{equation*}
\end{lemma}

The proof of this lemma and some properties of the operators in the Hilbert space are given in the appendix. 

\begin{remark}
The constant $M_0$ is related to the upper bound $M$ of $k(\mathbf{x}, \mathbf{x})$. 
It can be considered as a measure of richness of the feature space. 
The larger $M_0$ or $M$ is, the more complex the space is. 
\end{remark}

\begin{remark}
There are three terms in the error bound. 
The first term is of order $\mathcal{O}\left ( \frac{1}{\sqrt{t}} \right )$, which results from the estimation error of $\widehat{\boldsymbol{\Sigma}}$ and $\widehat{\boldsymbol{\Gamma}}$. 
The second term $\mathcal{O}\left ( \frac{1}{t} \right )$ comes from the construction of the dictionary. 
Since the cardinality of dictionary samples is finite, this error is bounded when the dictionary reaches its final size and will descend in order $\frac{1}{t}$ when $t$ tends to infinity. 
The third one $M_0 \sqrt{\nu} + \nu$ is controlled by the threshold parameter $\nu$ and the constant $M_0$ which is associated with the richness of the feature space. 
\end{remark}

Lemma \ref{lem2} states the concentration property of $\widetilde{\boldsymbol{\Sigma}}_t$ and $\widetilde{\boldsymbol{\Gamma}}_t$. 
Based on this we have the concentration inequality for $\widetilde{\mathbf{T}}_t = \widetilde{\boldsymbol{\Sigma}}_t^{-1} \widetilde{\boldsymbol{\Gamma}}_t$. 


\begin{theorem}
\label{thm2}
Under Assumption \ref{assum1} and Condition \ref{cond1}, $\forall \epsilon > 0$, there exists $T > 0$, for some constants $\delta$, $C_1$, $C_2$, $M_0 > 0$, for $t > T$, we have 
\begin{equation*}
\mathrm{Pr} \left(\left\|\widetilde{\mathbf{T}}_t - \mathbf{T}\right\| > \frac{\delta}{\sqrt{t}} + \frac{C_1}{t} + C_2 ( M_0 \sqrt{\nu} + \nu ) \right) < \epsilon . 
\end{equation*}
\end{theorem}

The proof of this theorem is given in the Appendix. 
Lemma \ref{lem2} and Theorem \ref{thm2} shows that the gaps between our reduced-order estimations and the true operators are bounded in probability. 
Based on Lemma 1 in Ferr\'e and Yao~\cite{Ferre2003fsir}, we can draw the following corollary on the gap between eigenvectors of $\widetilde{\mathbf{T}}_t$ and $\mathbf{T}$. 

\begin{corollary}
\label{coro1}
Under the conditions of Theorem \ref{thm2}, suppose $\mathbf{T}$ has distinct eigenvalues, then $\forall \epsilon > 0$, there exists $T > 0$, for some constants $\delta$, $C_1$, $C_2$, $M_0 > 0$, for $t > T$, the following holds
\begin{equation*}
\mathrm{Pr} \left(\left|\left\langle\widetilde{\boldsymbol{\beta}}_{t,j}, \phi(\cdot)\right\rangle-\left\langle\boldsymbol{\beta}_j, \phi(\cdot)\right\rangle\right| > \frac{\delta}{\sqrt{t}} + \frac{C_1}{t} + C_2 ( M_0 \sqrt{\nu} + \nu ) \right) < \epsilon , 
\end{equation*}
where $\widetilde{\boldsymbol{\beta}}_{t,j}$ and $\boldsymbol{\beta}_j$ are the $j$th eigenvectors of $\widetilde{\mathbf{T}}_t$ and $\mathbf{T}$. 
\end{corollary}

\begin{remark}
Corollary \ref{coro1} states that the spectrum of $\widetilde{\mathbf{T}}_t$ is close to $\mathbf{T}$ with large probability, where $\widetilde{\mathbf{T}}_t$ is the reduced-order stochastic estimation of $\mathbf{T}$ at $t$ time step. 
Our algorithm is a stochastic optimization of this 
eigen-decomposition problem. 
The convergence rate of this algorithm is given in Chen et al.~\cite{chen2019online}. 
\end{remark}

\subsection{Centering the Data in Feature Space}

In the previous sections, we have assumed that the mapped data are centered in the feature space, which means $\sum_{i=1}^{t} \phi(\mathbf{x}_i) = 0$. 
In practice, the kernel function maps the data into, almost always, an uncentered embedding. 
Wu~\cite{WuHM} showed that the process of centering the data in the feature space can be done directly on the kernel matrix without explicitly computing the mapping 
$$
\mathbf{K}^{c} = \mathbf{K}-\frac{1}{n} \mathbf{1}_n \mathbf{1}_n^{\top} \mathbf{K}-\frac{1}{n} \mathbf{K} \mathbf{1}_n \mathbf{1}_n^{\top}+\frac{1}{n^2} \mathbf{1}_n \mathbf{1}_n^{\top} \mathbf{K} \mathbf{1}_n \mathbf{1}_n^{\top} , 
$$
where $\mathbf{1}_n$ is a column vector of length $n$ composed of $1$. 
Thus, the kernel vector for a sample $\mathbf{x}$ is centralized as 
$$
\mathbf{k}(\mathbf{x})^{c} = \mathbf{k}(\mathbf{x}) - \frac{1}{n} \mathbf{1}_n \mathbf{1}_n^{\top} \mathbf{k}(\mathbf{x}) - \frac{1}{n} \mathbf{K} \mathbf{1}_n + \frac{1}{n^2} \mathbf{1}_n \mathbf{1}_n^{\top} \mathbf{K} \mathbf{1}_n . 
$$

For the proposed reduced-order method, Honeine~\cite{honeine2011online} states that the center is tracked in the subspace, which is estimated as  
$\bar{\mathbf{a}}_t = \frac{1}{t} \sum_{i=1}^{t} \mathbf{a}_i$ that can be updated recursively in our algorithm. 
This leads to an expression similar to the offline centralization 
$$
\widetilde{\mathbf{K}}_{t}^{c}=\widetilde{\mathbf{K}}_{t}-\mathbf{1}_{m_{t}} \bar{\mathbf{a}}_{t}^{\top} \widetilde{\mathbf{K}}_{t}-\widetilde{\mathbf{K}}_{t} \bar{\mathbf{a}}_{t} \mathbf{1}_{m_{t}}^{\top}+\mathbf{1}_{m_{t}} \bar{\mathbf{a}}_{t}^{\top} \widetilde{\mathbf{K}}_{t} \bar{\mathbf{a}}_{t} \mathbf{1}_{m_{t}}^{\top} , 
$$
and the kernel vector is centralized as 
$$
\widetilde{\mathbf{k}}_{t}^{c}(\mathbf{x})=\widetilde{\mathbf{k}}_{t}(\mathbf{x})-\mathbf{1}_{m_{t}} \bar{\mathbf{a}}_{t}^{\top} \widetilde{\mathbf{k}}_{t}(\mathbf{x})-\widetilde{\mathbf{K}}_{t} \bar{\mathbf{a}}_{t}+\mathbf{1}_{m_{t}} \bar{\mathbf{a}}_{t}^{\top} \widetilde{\mathbf{K}}_{t} \bar{\mathbf{a}}_{t} . 
$$
The centralized kernel matrix and kernel vector can be directly implemented into Algorithm \ref{alg1}, resulting in a centralized version of the algorithm. 
At $t$ step, the centralized algorithm yields eigenvectors $\boldsymbol{\Phi}_t^c = (\boldsymbol{\alpha}_{t,1}^{c}, \boldsymbol{\alpha}_{t,2}^{c}, \ldots, \boldsymbol{\alpha}_{t,d}^{c})$. 
The resulting model output is $\hat{v}_{t,j} = \hat{\boldsymbol{\alpha}}_{t,j}^{c \top} \widetilde{\mathbf{k}}_t^c(\mathbf{x})$. 

\section{Numerical Studies}

\label{sec4}

\subsection{Simulation Studies}

To investigate the effect of the online kernel sliced inverse regression algorithm, we tested our proposed algorithm in both linear and nonlinear settings and compared it with several other online dimension reduction methods as well as the batch kernel sliced inverse regression method. 
Several methods used for comparison include gradient descent based online SIR~\cite{cai2020online}, perturbation based online SIR~\cite{cai2020online}, online incremental PCA~\cite{hall1998incremental}, and reduced-order online KPCA~\cite{honeine2011online}, which we denote as OSIR(GD), OSIR(P), OPCA, OKPCA, respectively. 

First, we consider a generalized linear model~\cite{Li1991SIR}. 
We generate independent variable $\mathbf{x}$ from the multivariate normal distribution $N_{p}\left(\mathbf{0}, \boldsymbol{\Sigma}_{\mathbf{x}}\right)$, where the $i, j$-th element of the covariance matrix is $\left(\boldsymbol{\Sigma}_{\mathbf{x}}\right)_{ i, j}=0.5^{\lvert i-j \rvert}$. 
The random error $\varepsilon$ is generated by the normal distribution $N(0,1)$. 
The response variable $y$ satisfies:
$$
y=\frac{x_{1}+x_{2}+x_{3}}{0.5+\left(x_{4}+x_{5}+1.5\right)^{2}}+\varepsilon . 
$$
In this model, the e.d.r. directions are linear, as $v_{1}=\left\langle\boldsymbol{\beta}_{1}, \phi(\mathbf{x})\right\rangle=x_{1}+x_{2}+x_{3}$ and $v_{2}=\left\langle\boldsymbol{\beta}_{2}, \phi(\mathbf{x})\right\rangle=x_{4}+x_{5}$, where the mapping $\phi(\mathbf{x}) = \mathbf{x}$, and the dimension of the e.d.r. space $d=2$. 
We use the above model to generate the data stream and obtain the e.d.r. directions by different methods. 
The kernel function for OKSIR and OKPCA is chosen as the following additive Gaussian kernel:
$$
k(\mathbf{x}, \mathbf{z})=\sum_{j=1}^{p} \exp \left(-\frac{\left(x_{j}-z_{j}\right)^{2}}{2 \sigma^{2}}\right) . 
$$
We take the window width $\sigma$ to be 2. 
We repeat our simulations $N = 100$ times with sample size $n = 1000, 2000, 4000$ and covariate dimension $p = 100, 200, 400, 1000$. 
We also generate $n = 1000$ samples in the same way as the test set. 
Then we can get the estimated summary statistic $\hat{v}$ on the test set. 
To evaluate the performance of different methods, we consider the absolute correlation coefficients between $\hat{v}$'s and $v$'s. 
Since the positive and negative of the statistics may not be the same, here we refer the absolute correlation coefficients to the absolute value of the correlation coefficients. 

The results are shown in Table~\ref{tab1}, where the results of OSIR based on perturbation involve the inverse operation of a high dimension matrix and cannot give stable results when $p=1000$. 
As we can see, supervised dimension reduction methods perform significantly better than unsupervised methods in this setting, which is caused by the model setting. 
With the increase in data volume, the effectiveness of both OSIR methods and our OKSIR method improves. 
With the increase in covariate dimension, the performance of the OSIR method based on perturbation decreases a lot, and the performance of the OSIR method based on gradient descent also decreases. 
When the dimensionality is high, both OSIR methods lose their effectiveness, and our method remains robust. 
Overall, from the perspective of the correlation coefficient, our method achieves better dimension reduction performance than the other four methods. 

\begin{table}
\begin{center}
\begin{minipage}{\textwidth}
\centering
\caption{\small The averages of the absolute correlation coefficients between the estimated and true value of the sufficient dimension reduction variables based on 100 simulations, in brackets are the standard deviations. }
\label{tab1}
{\small
\begin{tabular}{cccccccc}
\toprule
               &      & OKSIR      & OSIR(GD)   & OSIR(P)    & OKPCA      & OPCA       \\
\hline p=100, n=1000  & cor1 & 0.66(0.05) & 0.61(0.10) & 0.62(0.09) & 0.19(0.09) & 0.17(0.09) \\
               & cor2 & 0.55(0.06) & 0.49(0.09) & 0.50(0.09) & 0.20(0.10) & 0.22(0.11) \\
p=100, n=2000  & cor1 & 0.70(0.04) & 0.64(0.08) & 0.67(0.08) & 0.17(0.09) & 0.15(0.07) \\
               & cor2 & 0.58(0.05) & 0.49(0.09) & 0.52(0.08) & 0.19(0.08) & 0.19(0.09) \\
p=100, n=4000  & cor1 & 0.72(0.04) & 0.64(0.09) & 0.71(0.07) & 0.17(0.09) & 0.15(0.09) \\
               & cor2 & 0.59(0.04) & 0.50(0.09) & 0.56(0.07) & 0.17(0.09) & 0.21(0.11) \\
p=200, n=1000  & cor1 & 0.60(0.04) & 0.49(0.14) & 0.25(0.17) & 0.13(0.06) & 0.12(0.07) \\
               & cor2 & 0.47(0.07) & 0.44(0.11) & 0.21(0.15) & 0.13(0.07) & 0.14(0.08) \\
p=200, n=2000  & cor1 & 0.64(0.04) & 0.48(0.13) & 0.26(0.18) & 0.12(0.06) & 0.11(0.06) \\
               & cor2 & 0.51(0.05) & 0.41(0.12) & 0.25(0.15) & 0.13(0.07) & 0.15(0.07) \\
p=200, n=4000  & cor1 & 0.67(0.04) & 0.50(0.13) & 0.22(0.16) & 0.13(0.07) & 0.10(0.06) \\
               & cor2 & 0.55(0.05) & 0.43(0.14) & 0.21(0.15) & 0.14(0.07) & 0.13(0.08) \\
p=400, n=1000  & cor1 & 0.57(0.04) & 0.38(0.08) & 0.21(0.09) & 0.09(0.04) & 0.09(0.05) \\
               & cor2 & 0.43(0.05) & 0.29(0.07) & 0.17(0.08) & 0.09(0.05) & 0.10(0.05) \\
p=400, n=2000  & cor1 & 0.63(0.03) & 0.39(0.08) & 0.21(0.08) & 0.11(0.04) & 0.09(0.05) \\
               & cor2 & 0.50(0.04) & 0.28(0.07) & 0.18(0.08) & 0.10(0.05) & 0.11(0.06) \\
p=400, n=4000  & cor1 & 0.66(0.03) & 0.39(0.07) & 0.23(0.08) & 0.09(0.05) & 0.08(0.05) \\
               & cor2 & 0.53(0.04) & 0.28(0.07) & 0.18(0.07) & 0.10(0.05) & 0.11(0.05) \\
p=1000, n=1000 & cor1 & 0.48(0.04) & 0.27(0.07) & -          & 0.07(0.03) & 0.07(0.03) \\
               & cor2 & 0.36(0.05) & 0.20(0.06) & -          & 0.07(0.04) & 0.07(0.04) \\
p=1000, n=2000 & cor1 & 0.55(0.03) & 0.27(0.05) & -          & 0.07(0.04) & 0.06(0.03) \\
               & cor2 & 0.41(0.04) & 0.19(0.06) & -          & 0.06(0.03) & 0.07(0.03) \\
p=1000, n=4000 & cor1 & 0.60(0.03) & 0.27(0.06) & -          & 0.07(0.03) & 0.06(0.04) \\
               & cor2 & 0.47(0.04) & 0.18(0.06) & -          & 0.07(0.03) & 0.08(0.04) \\\bottomrule
\end{tabular}
}
\end{minipage}
\end{center}
\end{table}

In addition, we also compare the running time of the algorithms under this simulation setting, and the results are shown in Table~\ref{tab2}. 
It can be seen that the computational efficiency of the incremental-based OPCA is the highest since one update of incremental-based OPCA costs $\mathcal{O}(p d^2)$ computations. 
The computation efficiency of our method is slightly better than the online SIR based on gradient descent and the reduced-order online KPCA, and significantly better than the online SIR based on the perturbation method. 
In the vertical view, the running time of our method increases almost linearly with the increase of data volume, which is consistent with the time complexity of our algorithm. 
This also verifies Lemma~\ref{lem1} that the dictionary size has an upper bound under certain conditions. 

\begin{table}
\begin{center}
\begin{minipage}{\textwidth}
\centering
\caption{\small The averages of the computation time (in seconds) based on 100 replications for different methods}
\label{tab2}
{\small
\begin{tabular}{cccccccc}
\toprule
               & OKSIR  & OSIR(GD) & OSIR(P) & OKPCA  & OPCA  \\
\hline p=100, n=1000  & 0.97   & 1.05     & 11.05   & 1.3    & 0.28  \\
p=100, n=2000  & 2.09   & 2.2      & 24.54   & 2.78   & 0.59  \\
p=100, n=4000  & 4.54   & 4.91     & 53.56   & 5.83   & 1.21  \\
p=200, n=1000  & 1.96   & 2.34     & 70.43   & 3.29   & 0.37  \\
p=200, n=2000  & 4.17   & 5.05     & 152.49  & 6.91   & 0.73  \\
p=200, n=4000  & 9.17   & 10.69    & 324.37  & 14.72  & 1.53  \\
p=400, n=1000  & 9.98   & 12.21    & 303.8   & 13.43  & 1.32  \\
p=400, n=2000  & 23.17  & 27.03    & 641.15  & 30.35  & 2.87  \\
p=400, n=4000  & 47.97  & 53.73    & 1366.29 & 58.93  & 5.96  \\
p=1000, n=1000 & 47.01  & 86.18    & -       & 51.62  & 9.54  \\
p=1000, n=2000 & 120.61 & 186.26   & -       & 154.23 & 18.9  \\
p=1000, n=4000 & 268.55 & 382.73   & -       & 307.53 & 42.98 \\\bottomrule
\end{tabular}}
\end{minipage}
\end{center}
\end{table}

The second one we consider is a nonlinear model. 
We generate independent variable $\mathbf{x}$ from the multivariate normal distribution $N_{p}\left(\mathbf{0}, \mathbf{I} \right)$. 
The response variable $y$ comes from the following model:
$$
y=\left(\sin \left(x_{1}\right)+\sin \left(x_{2}\right)\right)\left(1+\sin \left(x_{3}\right)\right)+0.1 \varepsilon , 
$$
where the random error $\varepsilon$ is sampled from $N(0,1)$. 
The e.d.r. directions of this model are nonlinear, and the summary statistics are $\nu_{1}=\left\langle\boldsymbol{\beta}_{1}, \phi(\mathbf{x})\right\rangle=\sin \left(x_{1}\right)+\sin \left(x_{2}\right)$ and $v_{2}=\left\langle\boldsymbol{\beta}_{2}, \phi(\mathbf{x})\right\rangle= 1+ \sin \left(x_{3}\right)$. 
Similarly, we generate the data stream of the above model with 1000 test samples. 
The kernel function is also the additive Gaussian kernel. 
To evaluate the performance of dimension reduction, we conduct Gaussian kernel regression on the estimated summary statistics $\hat{v}$'s, where the window widths are determined by cross-validation. 
We repeat our simulations $N = 100$ times with sample size $n = 500, 1000, 2000$ and covariate dimension $p = 10, 20$. 

The results are shown in Table~\ref{tab3}, where we use the 5-fold cross-validation error on the test set as a measure. 
In this experiment, we add the batch kernel sliced inverse regression (batch KSIR) as the baseline method for comparison. 
Our method achieves the best results among all online dimension reduction methods and achieves results close to those of the batch KSIR method. 

\begin{table}
\begin{center}
\begin{minipage}{\textwidth}
\centering
\caption{\small The averages of the 5-fold cross-validation errors based on 100 replications under different settings, in brackets are the standard deviations. }
\label{tab3}
{\small
\begin{tabular}{cccccccc}
\toprule
 & OKSIR & OSIR(GD) & OSIR(P) & OKPCA & OPCA & batch KSIR \\
\hline n=500, p=10 & 0.32(0.09) & 0.60(0.12) & 0.60(0.12) & 1.10(0.13) & 1.09(0.14) & 0.20(0.06) \\
n=500, p=20 & 0.41(0.08) & 0.81(0.11) & 0.81(0.10) & 1.18(0.09) & 1.18(0.09) & 0.32(0.06) \\
n=1000, p=10 & 0.27(0.06) & 0.60(0.12) & 0.60(0.12) & 1.09(0.15) & 1.09(0.13) & 0.14(0.04) \\
n=1000, p=20 & 0.33(0.06) & 0.80(0.11) & 0.80(0.11) & 1.18(0.09) & 1.18(0.09) & 0.21(0.03) \\
n=2000, p=10 & 0.24(0.06) & 0.58(0.10) & 0.58(0.10) & 1.10(0.13) & 1.10(0.16) & 0.11(0.02) \\
n=2000, p=20 & 0.29(0.05) & 0.80(0.10) & 0.80(0.10) & 1.18(0.09) & 1.18(0.11) & 0.15(0.02) \\\bottomrule
 \end{tabular}}
\end{minipage}
\end{center}
\end{table}

\subsection{Real Data Analysis}

To further compare the numerical performance of projection directions found by different approaches, we apply OKSIR, OSIR, OKPCA, and OPCA as feature extractors on real-world datasets. 
A brief description of these datasets is provided in Table~\ref{tab4}. 
The first eight datasets are from the UCI machine learning database~\cite{asuncion2007uci} and the last two are from the LIBSVM database~\cite{chang2011libsvm}. 
We randomly select 75\% of the samples as the training set and the rest as the test set. 
The kernel function is still the additive Gaussian kernel. 
After applying the dimension reduction methods to the dataset, we use the Support Vector Machine~(SVM) model in R package \verb|e1071| to construct the classifier or regressor. 
For datasets with a training sample size larger than 1000, we randomly select 1000 samples as the basis when implementing batch KSIR. 
We use the error rate $\sum_{i \in \text {testset}} I \left(y_{i} \neq \hat{y}_{i}\right) / \text{ntest}$ as the evaluation standards for categorical data. 
For regression task, we use the relative prediction error $\sum_{i \in \text { testset }}\left(y_{i}-\hat{y}_{i}\right)^{2} / \sum_{i \in \text { testset }}\left(y_{i}-\bar{y}\right)^{2}$ as the measurement. 
In our experiments, in addition to the online dimension reduction methods and the batch KSIR method, we also take into account the SVM model directly using the original features, denoted as Origin. 

\begin{table}
\begin{center}
\begin{minipage}{\textwidth}
\centering
\caption{\small The description of the datasets}
\label{tab4}
{\small
\begin{tabular}{cccccccc}
\toprule
dataset & full name & task & sample size & p & d \\
\hline WBC & Wisconsin breast cancer & classification & 699 & 9 & 1 \\
ION & Pima Indians diabetes & classification & 351 & 34 & 1 \\
PID & Pima Indians diabetes &classification & 768 & 8 & 1 \\
WAV & Waveform database generator & classification & 5000 & 40 & 2 \\
OPT & Optical recognition digits & classification & 5620 & 64 & 6 \\
ACT & Activity recognition & classification & 4480 & 533 & 3 \\
COM & Communities and crime & regression & 1994 & 100 & 4 \\
CT & Relative location of CT slices & regression & 53500 & 384 & 4 \\
MG & Mackey-Glass delay-differential equation & regression & 1385 & 6 & 2 \\
CPU & Computer activity & regression & 8192 & 12 & 3 \\\bottomrule
 \end{tabular}}
\end{minipage}
\end{center}
\end{table}

The above procedure was repeated 100 times randomly and the results are shown in Table \ref{tab5}. 
We can see that the OKSIR method outperforms other online dimension reduction methods on real-world datasets, consistent with previous research work on kernel sliced inverse regression. 
Compared with using the original data directly, our method performs better in the high-dimensional case, which indicates that our approach can accurately capture the low-dimensional intrinsic structure of high-dimensional data. 
Moreover, the performance of our proposed method is similar to that of the batch KSIR. 
That means our streaming model can achieve a level close to the batch processing model. 

\begin{table}
\begin{center}
\begin{minipage}{\textwidth}
\centering
\caption{\small The average performances of different methods on real datasets based on 100 replications, in brackets are the standard deviations. }
\label{tab5}
{\small 
\begin{tabular}{cccccccc}
\toprule
    & OKSIR & OSIR(GD) & OSIR(P) & OKPCA & OPCA & batch KSIR & Origin \\\hline
WBC & 0.031 & 0.053 & 0.034 & 0.039 & 0.032 & 0.030 & 0.035 \\
& (0.010) & (0.014) & (0.011) & (0.013) & (0.011) & (0.010) & (0.011) \\
ION & 0.106 & 0.250 & 0.366 & 0.323 & 0.371 & 0.105 & 0.135 \\
& (0.030) & (0.044) & (0.045) & (0.080) & (0.046) & (0.030) & (0.030) \\
PID & 0.250 & 0.346 & 0.347 & 0.326 & 0.348 & 0.262 & 0.270 \\
& (0.028) & (0.032) & (0.030) & (0.043) & (0.028) & (0.026) & (0.027) \\
WAV & 0.135 & 0.240 & 0.136 & 0.268 & 0.275 & 0.134 & 0.140 \\
& (0.009) & (0.019) & (0.009) & (0.070) & (0.011) & (0.009) & (0.009) \\
OPT & 0.058 & 0.113 & 0.136 & 0.150 & 0.073 & 0.060 & 0.026 \\
& (0.007) & (0.024) & (0.042) & (0.026) & (0.012) & (0.008) & (0.004) \\
ACT & 0.348 & 0.598 & 0.539 & 0.443 & 0.552 & 0.341 & 0.763 \\
& (0.014) & (0.027) & (0.033) & (0.042) & (0.017) & (0.018) & (0.005) \\
COM & 0.410 & 0.489 & 0.499 & 0.472 & 0.440 & 0.408 & 0.458 \\
& (0.031) & (0.036) & (0.056) & (0.065) & (0.041) & (0.031) & (0.031) \\
CT & 0.144 & 0.147 & 0.153 & 0.379 & 0.230 & 0.289 & 0.216 \\
& (0.034) & (0.036) & (0.031) & (0.097) & (0.022) & (0.027) & (0.005) \\
MG & 0.375 & 0.386 & 0.522 & 0.560 & 0.758 & 0.372 & 0.293 \\
& (0.029) & (0.030) & (0.184) & (0.155) & (0.051) & (0.029) & (0.026) \\
CPU & 0.037 & 0.411 & 0.359 & 0.176 & 0.033 & 0.055 & 0.045 \\
& (0.005) & (0.115) & (0.289) & (0.192) & (0.003) & (0.022) & (0.006) \\\hline 
\end{tabular}}
\end{minipage}
\end{center}
\end{table}

In our previous analysis, we built up a prediction model using SVM. 
In the following, we compare SVM with classification tree, linear discriminant analysis, generalized linear model and random forest to build up prediction models. 
We use "WAV" data as an example. 
The prediction errors are summarized in Table \ref{tab6}. 
All of these numbers show similar patterns. 
This suggests that using SVM to build up predictive models does not have any significant impact in our comparative study. 

\begin{table}
\begin{center}
\begin{minipage}{\textwidth}
\centering
\caption{\small The prediction errors based on the “WAV” data. The models are built up with support vector machines (SVM), classification tree (TREE), linear discriminant analysis (LDA), generalized linear model (GLM), and random forest (RF), in brackets are the standard deviations. }
\label{tab6}
{\small 
\begin{tabular}{cccccccc}
\toprule
    & OKSIR & OSIR(GD) & OSIR(P) & OKPCA & OPCA & batch KSIR & Origin \\\hline
SVM & 0.135 & 0.240 & 0.136 & 0.268 & 0.275 & 0.134 & 0.140 \\
& (0.009) & (0.019) & (0.009) & (0.070) & (0.011) & (0.009) & (0.009) \\
TREE & 0.163 & 0.223 & 0.164 & 0.299 & 0.289 & 0.161 & 0.263 \\
& (0.011) & (0.019) & (0.011) & (0.065) & (0.011) & (0.011) & (0.011) \\
LDA & 0.143 & 0.206 & 0.144 & 0.274 & 0.279 & 0.141 & 0.142 \\
& (0.010) & (0.017) & (0.010) & (0.072) & (0.011) & (0.010) & (0.009) \\
GLM & 0.136 & 0.204 & 0.139 & 0.285 & 0.282 & 0.135 & 0.133 \\
& (0.009) & (0.017) & (0.009) & (0.080) & (0.010) & (0.010) & (0.010) \\
RF & 0.152 & 0.227 & 0.157 & 0.309 & 0.312 & 0.149 & 0.145 \\
& (0.010) & (0.020) & (0.009) & (0.084) & (0.012) & (0.009) & (0.009) \\
\hline 
\end{tabular}}
\end{minipage}
\end{center}
\end{table}

\section{Conclusions}

\label{sec5}

In this paper, we have proposed an online kernel sliced inverse regression method, which achieves similar performance to the batch kernel sliced inverse regression method. 
This online fashion consists of two steps. 
One is the introduction of ALD condition and dictionary samples, which help us update the variables in the original problem online. 
Another is the online update for the eigenvectors of (\ref{eq5}). 
We also give an approach to online centering the data in the feature space. 
Numerical studies show that our method successfully implements online nonlinear dimension reduction and can accurately extract both linear and nonlinear e.d.r. directions. 
Our method can be extended to other nonlinear dimension reduction methods, such as the nonlinear sufficient dimension reduction method for functional data~\cite{li2017fsdr}. 

\section*{Appendix A}

\subsection*{A.1 Preliminaries}

In this section, we briefly introduce some basic concepts of operators in Hilbert space. 
A linear operator (or simply an operator) $\mathbf{L}$ is defined on $\mathcal{H}$ to $\mathcal{H}'$ as a function which sends every vector $u$ in $\mathcal{H}$ to a vector $v = \mathbf{L} u$ in $\mathcal{H}'$ and satisfies the linearity condition that $$\mathbf{L}(\alpha_1 u_1 + \alpha_2 u_2) = \alpha_1 \mathbf{L} u_1 + \alpha_2 \mathbf{L} u_2 ,  $$for all $u_1$, $u_2$ in $\mathcal{H}$ and $\alpha_1$, $\alpha_2 \in \mathcal{R}$. 
An operator $\mathbf{L}$ is bounded if $\| \mathbf{L} u \| \leq M \|u\|$, and $\mathbf{L}$ is continuous if and only if $\mathbf{L}$ is bounded. 
$\|\cdot\|$ denotes the operator norm $$\|\mathbf{L}\| = \mathop{sup}_{u \neq 0} \frac{\|\mathbf{L} u\|}{\|u\|} . $$
An operator $\mathbf{L}$ is compact if the image $\{\mathbf{L} u_i \}$ of any bounded sequence $\{u_i \}$ of $\mathcal{H}$ contains a Cauchy subsequence. 

For any subset $\mathcal{S}$ of $\mathcal{H}$, the set of all vectors of the form $\mathbf{L} u$ with $u \in \mathcal{S}$ is called the image under $\mathbf{L}$ of $\mathcal{S}$ and is denoted by $\mathbf{L} \mathcal{S}$. 
In particular, the linear manifold $\mathbf{L} \mathcal{H}$ of $\mathcal{H}'$ is called the range of $\mathbf{L}$ and is denoted by $\mathbf{R}(\mathbf{L})$. 
The dimension of $\mathbf{R}(\mathbf{L})$ is called the rank of $\mathbf{L}$. 
An operator $\mathbf{L}$ is said to be degenerate if $\text{rank}(\mathbf{L})$ is finite, and a degenerate operator is compact. 
The adjoint operator of $\mathbf{L}$ is denoted as $\mathbf{L}^*$ and satisfies $\langle \mathbf{L} u, v \rangle = \langle u, \mathbf{L}^* v \rangle$. 
An operator $\mathbf{L}$ is said to be symmetric if $\mathbf{L} \subset \mathbf{L}^*$ and is self-adjoint if $\mathbf{L} = \mathbf{L}^*$. 
A complex number $\lambda$ is an eigenvalue of $\mathbf{L}$ if there is a non-zero vector $u \in \mathcal{H}$ such that $\mathbf{L} u = \lambda u$, where $u$ is an eigenvector. 
The set of all eigenvalues of $\mathbf{L}$ is called the spectrum of $\mathbf{L}$, denoted as $\sigma(\mathbf{L})$. 

One of the most important classes of compact operators is the Hilbert-Schmidt class. 
Define Hilbert-Schmidt norm of $\mathbf{L}$ as $$\|\mathbf{L}\|_{\mathrm{HS}} = (\sum_{i=1}^{\infty} \|\mathbf{L}\varphi_i\|^2)^{1/2} , $$where $\{\varphi_i\}$ is a complete orthonormal family in $\mathcal{H}$. 
The set consisting of all $\mathbf{L}$ with $\|\mathbf{L}\|_{HS} < \infty$ is called the Hilbert-Schmidt class. 
Given $\mathbf{L}$ belongs to the Hilbert-Schmidt class and $\mathbf{S}$ is a bounded operator, $\mathbf{L S}$ and $\mathbf{S L}$ belongs to the Hilbert-Schmidt class and we have the inequalities $$\|\mathbf{L S}\|_{\mathrm{HS}} \leq \|\mathbf{L}\|_{\mathrm{HS}} \|\mathbf{S}\| , \quad \|\mathbf{S L}\|_{\mathrm{HS}} \leq \|\mathbf{S}\| \|\mathbf{L}\|_{\mathrm{HS}} , \quad \|\mathbf{L}\| \leq \|\mathbf{L}\|_{\mathrm{HS}} . $$
We can introduce an inner product for the Hilbert-Schmidt class so that it becomes a Hilbert space. 
We set $$\langle \mathbf{L}, \mathbf{S} \rangle = \sum_{i=1}^{\infty} \langle \mathbf{L} \varphi_i, \mathbf{S} \varphi_i \rangle , $$where $\mathbf{L}$ and $\mathbf{S}$ belong to the Hilbert-Schmidt class. 
For more details on the operator theory, see Kato's book \cite{kato2013perturbation} and the references therein. 

\subsection*{A.2 Proof of Lemma \ref{lem2}}

We first consider the concentration inequality of $\widetilde{\boldsymbol{\Sigma}}$, and the following holds $$\left \|\widetilde{\boldsymbol{\Sigma}}_t - \boldsymbol{\Sigma}_t\right\|_{\mathrm{HS}} \leq \left \|\widetilde{\boldsymbol{\Sigma}}_t - \widehat{\boldsymbol{\Sigma}}_t\right\|_{\mathrm{HS}} + \left\|\widehat{\boldsymbol{\Sigma}}_t - \boldsymbol{\Sigma}_t\right\|_{\mathrm{HS}} . $$

For $\left\|\widetilde{\boldsymbol{\Sigma}}_t - \widehat{\boldsymbol{\Sigma}}_t\right\|_{\mathrm{HS}}$, we have
\begin{align*}
\left\|\widetilde{\boldsymbol{\Sigma}}_t - \widehat{\boldsymbol{\Sigma}}_t\right\|_{\mathrm{HS}} & = \left \| \frac{1}{t} \sum_{i=1}^{t} \widetilde{\phi}(\mathbf{x}_i) \otimes \widetilde{\phi}(\mathbf{x}_i) - \frac{1}{t} \sum_{i=1}^{t} \phi(\mathbf{x}_i) \otimes \phi(\mathbf{x}_i) \right \|_{\mathrm{HS}} \\
& = \left \| \frac{1}{t} \sum_{i=1}^{t} \left [ \left ( \phi(\mathbf{x}_i) + \phi_{i}^{res} \right ) \otimes \left ( \phi(\mathbf{x}_i) + \phi_{i}^{res} \right ) - \phi(\mathbf{x}_i) \otimes \phi(\mathbf{x}_i) \right ] \right \|_{\mathrm{HS}} \\
& \leq \frac{1}{t} \sum_{i=1}^{t} \left ( \left \| \phi_{i}^{res} \right \|^2 + 2 \left \| \phi(\mathbf{x}_i) \right \| \left \| \phi_{i}^{res} \right \| \right ) \\
& \leq \frac{1}{t} (C_1 + \nu t + M_0 \sqrt{\nu} t) . 
\end{align*}
The first inequality is directly from the definition of Hilbert-Schmidt norm and the Parseval equality. 
The second inequality results from the definition of the ALD condition and Proposition 2.3 and its discussion in Engel et al.~\cite{engel2004kernel}. 

We will use the following results from Ferr\'e and Yao~\cite{Ferre2003fsir}
\begin{equation*}
\left \|\widehat{\boldsymbol{\Sigma}}_t-\boldsymbol{\Sigma} \right \|_{\mathrm{HS}}=O_p\left(\frac{1}{\sqrt{t}}\right) , \quad \left \|\widehat{\boldsymbol{\Gamma}}_t-\boldsymbol{\Gamma} \right \|_{\mathrm{HS}}=O_p\left(\frac{1}{\sqrt{t}}\right) . 
\end{equation*}

Combining these results, $\forall \epsilon > 0$, there exists $\delta_1$, $T_1 > 0$, for $t > T_1$, we have 
\begin{equation*}
\mathrm{Pr} \left(\left\|\widetilde{\boldsymbol{\Sigma}}_t - \boldsymbol{\Sigma}\right\|_{HS} > \frac{\delta_1}{\sqrt{t}} + \frac{C_1}{t} + M_0 \sqrt{\nu} + \nu \right) < \epsilon . 
\end{equation*}

As for $\widetilde{\boldsymbol{\Gamma}}_t$, we have 
\begin{equation*}
\left \| \widetilde{\boldsymbol{\Gamma}}_t - \boldsymbol{\Gamma} \right \|_{\mathrm{HS}} \leq \left \| \widetilde{\boldsymbol{\Gamma}}_t - \widehat{\boldsymbol{\Gamma}}_t \right \|_{\mathrm{HS}} + \left \| \widehat{\boldsymbol{\Gamma}}_t - \boldsymbol{\Gamma} \right \|_{\mathrm{HS}}. 
\end{equation*}
Similarly, for $\left \| \widetilde{\boldsymbol{\Gamma}}_t - \widehat{\boldsymbol{\Gamma}}_t \right \|_{\mathrm{HS}}$, we have 
\begin{align*}
\left \| \widetilde{\boldsymbol{\Gamma}}_t - \widehat{\boldsymbol{\Gamma}}_t \right \|_{\mathrm{HS}} & = \left \| \sum_{h=1}^{H} \frac{n_{h,t}}{t} \left [ \left ( \frac{1}{n_{h,t}} \sum_{y_i \in I_h} \widetilde{\phi}(\mathbf{x}_i) \right ) \otimes \left ( \frac{1}{n_{h,t}} \sum_{y_i \in I_h} \widetilde{\phi}(\mathbf{x}_i) \right ) \right. \right. \\
& \left. \left. - \left ( \frac{1}{n_{h,t}} \sum_{y_i \in I_h} \phi(\mathbf{x}_i) \right ) \otimes \left ( \frac{1}{n_{h,t}} \sum_{y_i \in I_h} \phi(\mathbf{x}_i) \right ) \right ] \right \|_{\mathrm{HS}} \\
& = \left \| \frac{1}{t} \sum_{h=1}^{H} \frac{1}{n_{h,t}} \sum_{y_i, y_j \in I_h} \left ( \widetilde{\phi}(\mathbf{x}_i) \otimes \widetilde{\phi}(\mathbf{x}_j) - \phi(\mathbf{x}_i) \otimes \phi(\mathbf{x}_j) \right ) \right \|_{\mathrm{HS}} \\
& \leq \frac{1}{t} \sum_{h=1}^{H} \frac{1}{n_{h,t}} \sum_{y_i, y_j \in I_h} \left ( \left \| \phi_{i}^{res} \right \| \left \| \phi_{j}^{res} \right \| + \left \| \phi(\mathbf{x}_i) \right \| \left \| \phi_{j}^{res} \right \| + \left \| \phi(\mathbf{x}_j) \right \| \left \| \phi_{i}^{res} \right \| \right ) \\
& \leq \frac{1}{t} \left ( C_2 + \sum_{h=1}^{H} n_{h,t} \left ( \nu + M_0 \sqrt{\nu} \right ) \right ) \\
& = \frac{1}{t} \left ( C_2 + \nu t + M_0 \sqrt{\nu} t \right ) . 
\end{align*}

Also, combining these results, $\forall \epsilon > 0$, there exists $\delta_2$, $T_2 > 0$, for $t > T_2$, we have 
\begin{equation*}
\mathrm{Pr} \left(\left\|\widetilde{\boldsymbol{\Gamma}}_t - \boldsymbol{\Gamma}\right\|_{HS} > \frac{\delta_2}{\sqrt{t}} + \frac{C_2}{t} + M_0 \sqrt{\nu} + \nu \right) < \epsilon . 
\end{equation*}

Use the inequality that $\|\mathbf{L}\| \leq \|\mathbf{L}\|_{\mathrm{HS}}$, we complete the proof. 

\section*{A.3 Proof of Theorem \ref{thm2}}

We have 
\begin{align*}
\left \| \widetilde{\mathbf{T}}_t - \mathbf{T} \right \|_{\mathrm{HS}} & = \left \| \widetilde{\boldsymbol{\Sigma}}^{-1}_{t} \widetilde{\boldsymbol{\Gamma}}_{t} - \boldsymbol{\Sigma}^{-1} \boldsymbol{\Gamma} \right \|_{\mathrm{HS}} \\
& \leq \left \| \widetilde{\boldsymbol{\Sigma}}^{-1}_{t} \widetilde{\boldsymbol{\Gamma}}_{t} - \widetilde{\boldsymbol{\Sigma}}^{-1}_{t} \boldsymbol{\Gamma} \right \|_{\mathrm{HS}} + \left \| \widetilde{\boldsymbol{\Sigma}}^{-1}_{t} \boldsymbol{\Gamma} - \boldsymbol{\Sigma}^{-1} \boldsymbol{\Gamma} \right \|_{\mathrm{HS}} \\
& \leq \left \| \widetilde{\boldsymbol{\Sigma}}^{-1}_{t} \right \| \left \| \widetilde{\boldsymbol{\Gamma}}_{t} - \boldsymbol{\Gamma} \right \|_\mathrm{HS} + \left \| \left ( \widetilde{\boldsymbol{\Sigma}}^{-1}_{t} - \boldsymbol{\Sigma}^{-1} \right ) \boldsymbol{\Gamma} \right \|_{\mathrm{HS}} . 
\end{align*}

For the second term, note that
\begin{equation*}
\left ( \widetilde{\boldsymbol{\Sigma}}^{-1}_{t} - \boldsymbol{\Sigma}^{-1} \right ) \boldsymbol{\Gamma} = \sum_{i=1}^{d_{\Gamma}} \tau_i \left ( \widetilde{\boldsymbol{\Sigma}}^{-1}_{t} - \boldsymbol{\Sigma}^{-1} \right ) \varphi_i \otimes \varphi_i . 
\end{equation*}
According to the definition of Hilbert-Schmidt norm, we have 
\begin{align*}
\left \| \left ( \widetilde{\boldsymbol{\Sigma}}^{-1}_{t} - \boldsymbol{\Sigma}^{-1} \right ) \boldsymbol{\Gamma} \right \|_{\mathrm{HS}}^2 & = \sum_{j=1}^{d_{\Gamma}} \left \| \sum_{i=1}^{d_{\Gamma}} \tau_i \left ( \widetilde{\boldsymbol{\Sigma}}^{-1}_{t} - \boldsymbol{\Sigma}^{-1} \right ) \varphi_i \langle \varphi_i, \varphi_j \rangle \right \|^2 \\
& = \sum_{j=1}^{d_{\Gamma}} \left \| \tau_j \left ( \widetilde{\boldsymbol{\Sigma}}^{-1}_{t} - \boldsymbol{\Sigma}^{-1} \right ) \varphi_j \right \|^2 \\
& = \sum_{j=1}^{d_{\Gamma}} \left \| \tau_j \left ( \widetilde{\boldsymbol{\Sigma}}^{-1}_{t} - \boldsymbol{\Sigma}^{-1} \right ) \boldsymbol{\Sigma} \widetilde{\varphi}_j \right \|^2 \\
& = \sum_{j=1}^{d_{\Gamma}} \left \| \tau_j \widetilde{\boldsymbol{\Sigma}}^{-1}_{t} \left ( \boldsymbol{\Sigma} - \widetilde{\boldsymbol{\Sigma}}_{t} \right ) \widetilde{\varphi}_j \right \|^2 \\
& \leq \sum_{j=1}^{d_{\Gamma}} \tau_j^2 \left \| \widetilde{\boldsymbol{\Sigma}}^{-1}_{t} \right \|^2 \left \| \boldsymbol{\Sigma} - \widetilde{\boldsymbol{\Sigma}}_{t} \right \|_{\mathrm{HS}}^2 , 
\end{align*}
where the third equality results from (1) in Proposition \ref{prop1}. 
Thus, the following holds for some positive constant $C_{\Gamma}$
\begin{equation*}
\left \| \left ( \widetilde{\boldsymbol{\Sigma}}^{-1}_{t} - \boldsymbol{\Sigma}^{-1} \right ) \boldsymbol{\Gamma} \right \|_{\mathrm{HS}} \leq C_{\Gamma} \left \| \widetilde{\boldsymbol{\Sigma}}^{-1}_{t} \right \| \left \| \boldsymbol{\Sigma} - \widetilde{\boldsymbol{\Sigma}}_{t} \right \|_{\mathrm{HS}} . 
\end{equation*}

Combining the results, we have 
\begin{equation*}
\left \| \widetilde{\mathbf{T}}_t - \mathbf{T} \right \|_{\mathrm{HS}} \leq \left \| \widetilde{\boldsymbol{\Sigma}}^{-1}_{t} \right \| \left ( \left \| \widetilde{\boldsymbol{\Gamma}}_{t} - \boldsymbol{\Gamma} \right \|_\mathrm{HS} + C_{\Gamma} \left \| \widetilde{\boldsymbol{\Sigma}}_{t} - \boldsymbol{\Sigma} \right \|_{\mathrm{HS}} \right ) . 
\end{equation*}

From the construction of $\widetilde{\boldsymbol{\Sigma}}_{t}$, we can see that $\widetilde{\boldsymbol{\Sigma}}_{t}$ has finite rank when $t$ is sufficiently large in the online case. 
Thus $\widetilde{\boldsymbol{\Sigma}}_{t}$ is a self-adjoint and compact operator. 
Suppose $\mathop{rank}(\widetilde{\boldsymbol{\Sigma}}_{t}) = k$, we have the following representation
\begin{equation*}
\widetilde{\boldsymbol{\Sigma}}_{t} = \sum_{i=1}^{k} \widetilde{\omega}_i u_i \otimes u_i , 
\end{equation*}
where $\widetilde{\omega}_1 \geq \widetilde{\omega}_2 \geq \cdots \geq \widetilde{\omega}_k > 0$. 
Applying (i) of Lemma 1 in Ferr\'e and Yao~\cite{Ferre2003fsir}, we have 
\begin{equation*}
\widetilde{\omega}_k \geq \omega_k - \left \|\widetilde{\boldsymbol{\Sigma}}_{t} - \boldsymbol{\Sigma}\right \|_{\mathrm{HS}} , 
\end{equation*}
where $\omega_k$ is the corresponding eigenvalue of $\boldsymbol{\Sigma}$. 
There exists a constant $c > 0$ that $\omega_k \geq c^{-1}$. 
We have 
\begin{equation*}
\left \| \widetilde{\mathbf{T}}_t - \mathbf{T} \right \|_{\mathrm{HS}} \leq \frac{\left \| \widetilde{\boldsymbol{\Gamma}}_{t} - \boldsymbol{\Gamma} \right \|_{\mathrm{HS}} + C_{\Gamma} \left \| \widetilde{\boldsymbol{\Sigma}}_{t} - \boldsymbol{\Sigma} \right \|_{\mathrm{HS}}}{c^{-1} - \left \|\widetilde{\boldsymbol{\Sigma}}_{t} - \boldsymbol{\Sigma}\right \|_{\mathrm{HS}}} . 
\end{equation*}

When $\left \|\widetilde{\boldsymbol{\Sigma}}_{t} - \boldsymbol{\Sigma}\right \|_{\mathrm{HS}} \leq \frac{1}{2c}$, the following holds
\begin{equation*}
\left \| \widetilde{\mathbf{T}}_t - \mathbf{T} \right \|_{\mathrm{HS}} \leq 2 c \left ( \left \| \widetilde{\boldsymbol{\Gamma}}_{t} - \boldsymbol{\Gamma} \right \|_{\mathrm{HS}} + C_{\Gamma} \left \| \widetilde{\boldsymbol{\Sigma}}_{t} - \boldsymbol{\Sigma} \right \|_{\mathrm{HS}} \right ) . 
\end{equation*}
Based on the proof of Lemma \ref{lem2}, for all $\epsilon_1 > 0$, there exists $T_1$, $\delta$, $C_1$, $C_2 > 0$ that for $t > T_1$, we have
\begin{equation*}
\mathrm{Pr} \left ( \left \| \widetilde{\mathbf{T}}_t - \mathbf{T} \right \|_{\mathrm{HS}} > \frac{\delta}{\sqrt{t}} + \frac{C_1}{t} + C_2 (M_0 \sqrt{\nu} + \nu) \right ) < \epsilon . 
\end{equation*}
Notice that for all $\epsilon_2 > 0$, there exists $T_2 > 0$ that for $t > T_2$, the inequality $\left \|\widetilde{\boldsymbol{\Sigma}}_{t} - \boldsymbol{\Sigma}\right \|_{\mathrm{HS}} \leq \frac{1}{2c}$ holds with probability larger than $1 - \epsilon_2$, which completes the proof. 

\section*{Acknowledgments}

This work was supported by the National Natural Science Foundation of China (71873128). 

\section*{Conflict of Interest}

The authors declare that they have no conflict of interest. 

\section*{Biographies}

\begin{itemize}[leftmargin=*]
    \item[] \textbf{Wenquan Cui} received his Ph.D. degree in Statistics from the University of Science and Technology of China in 2004. 
He is currently an associate professor at the University of Science and Technology of China. 
His major research interests focus on survival analysis, high-dimensional statistical inference and statistical machine learning. 
    \item[] \textbf{Yue Zhao} received his Master degree in Statistics from the University of Science and Technology of China in 2022. 
His research interests focus on sufficient dimension reduction and statistical machine learning. 
    \item[] \textbf{Jianjun Xu} received his Ph.D. degree in Statistics from the University of Science and Technology of China in 2022.
His research interests focus on functional data analysis. 
    \item[] \textbf{Haoyang Cheng} received his Ph.D. degree in Statistics from the University of Science and Technology of China in 2021. 
He is currently a lecturer at Quzhou University. 
His major research interests focus on survival analysis and high-dimensional statistical inference. 
\end{itemize}


\bibliographystyle{oega}
\bibliography{sample.bib}

\begin{thebibliography}{27}
\providecommand{\natexlab}[1]{#1}

\bibitem[{Ajzerman et~al.(1964)Ajzerman, Braverman and Rozonoehr}]{Ajzerman}
Ajzerman, M., Braverman, E. and Rozonoehr, L. (1964).
\newblock Theoretical foundations of the potential function method in pattern
  recognition learning.
\newblock \textit{Automation and Remote Control} 25:821--837.

\bibitem[{Arora et~al.(2012)Arora, Cotter, Livescu and
  Srebro}]{arora2012stochastic}
Arora, R., Cotter, A., Livescu, K. and Srebro, N. (2012).
\newblock Stochastic optimization for pca and pls.
\newblock \textit{2012 50th Annual Allerton Conference on Communication,
  Control, and Computing (Allerton)} pp. 861--868.

\bibitem[{Asuncion and Newman(2007)}]{asuncion2007uci}
Asuncion, A. and Newman, D. (2007).
\newblock Uci machine learning repository.

\bibitem[{{Cai} et~al.(2020){Cai}, {Li} and {Zhu}}]{cai2020online}
{Cai}, Z., {Li}, R. and {Zhu}, L. (2020).
\newblock Online sufficient dimension reduction through sliced inverse
  regression.
\newblock \textit{Journal of Machine Learning Research} 21(10):1--25.

\bibitem[{Chang and Lin(2011)}]{chang2011libsvm}
Chang, C.C. and Lin, C.J. (2011).
\newblock Libsvm: a library for support vector machines.
\newblock \textit{ACM transactions on intelligent systems and technology
  (TIST)} 2(3):1--27.

\bibitem[{Chen et~al.(2019)Chen, Li, Yang, Haupt and Zhao}]{chen2019online}
Chen, Z., Li, X., Yang, L., Haupt, J. and Zhao, T. (2019).
\newblock On constrained nonconvex stochastic optimization: A case study for
  generalized eigenvalue decomposition.
\newblock In: K.~Chaudhuri and M.~Sugiyama (eds.) \textit{Proceedings of the
  Twenty-Second International Conference on Artificial Intelligence and
  Statistics}, \textit{Proceedings of Machine Learning Research}, vol.~89, pp.
  916--925. PMLR.

\bibitem[{Cook(2000)}]{Cook2000SAVE}
Cook, R.D. (2000).
\newblock Save: a method for dimension reduction and graphics in regression.
\newblock \textit{Communications in Statistics - Theory and Methods}
  29(9-10):2109--2121.

\bibitem[{Engel et~al.(2004)Engel, Mannor and Meir}]{engel2004kernel}
Engel, Y., Mannor, S. and Meir, R. (2004).
\newblock The kernel recursive least-squares algorithm.
\newblock \textit{IEEE Transactions on signal processing} 52(8):2275--2285.

\bibitem[{Ferr\'e and Yao(2003)}]{Ferre2003fsir}
Ferr\'e, L. and Yao, A.F. (2003).
\newblock Functional sliced inverse regression analysis.
\newblock \textit{Statistics} 37(6):475--488.

\bibitem[{F.R.S.(1901)}]{Pearson1901pca}
F.R.S., K.P. (1901).
\newblock Liii. on lines and planes of closest fit to systems of points in
  space.
\newblock \textit{The London, Edinburgh, and Dublin Philosophical Magazine and
  Journal of Science} 2(11):559--572.

\bibitem[{Ghashami et~al.(2016)Ghashami, Perry and
  Phillips}]{ghashami2016streaming}
Ghashami, M., Perry, D.J. and Phillips, J. (2016).
\newblock Streaming kernel principal component analysis.
\newblock In: \textit{Artificial intelligence and statistics}, pp. 1365--1374.
  PMLR.

\bibitem[{Hall et~al.(1998)Hall, Marshall and Martin}]{hall1998incremental}
Hall, P.M., Marshall, A.D. and Martin, R.R. (1998).
\newblock Incremental eigenanalysis for classification.
\newblock In: \textit{Proceedings of the British Machine Vision Conference},
  pp. 29.1--29.10. BMVA Press.

\bibitem[{Hastie et~al.(2009)Hastie, Tibshirani and Friedman}]{Hastie}
Hastie, T., Tibshirani, R. and Friedman, J. (2009).
\newblock \textit{The Elements of Statistical Learning: Data Mining, Inference,
  and Prediction, Second Edition (Springer Series in Statistics)}.
\newblock Springer New York, NY.

\bibitem[{Honeine(2011)}]{honeine2011online}
Honeine, P. (2011).
\newblock Online kernel principal component analysis: A reduced-order model.
\newblock \textit{IEEE transactions on pattern analysis and machine
  intelligence} 34(9):1814--1826.

\bibitem[{Kato(2013)}]{kato2013perturbation}
Kato, T. (2013).
\newblock \textit{Perturbation theory for linear operators}, vol. 132.
\newblock Springer Science \& Business Media.

\bibitem[{Kimura et~al.(2005)Kimura, Ozawa and Abe}]{kimura2005incremental}
Kimura, S., Ozawa, S. and Abe, S. (2005).
\newblock Incremental kernel pca for online learning of feature space.
\newblock \textit{International Conference on Computational Intelligence for
  Modelling, Control and Automation and International Conference on Intelligent
  Agents, Web Technologies and Internet Commerce (CIMCA-IAWTIC'06)} 1:595--600.

\bibitem[{Li and Song(2017)}]{li2017fsdr}
Li, B. and Song, J. (2017).
\newblock {Nonlinear sufficient dimension reduction for functional data}.
\newblock \textit{The Annals of Statistics} 45(3):1059 -- 1095.

\bibitem[{Li(1991)}]{Li1991SIR}
Li, K.C. (1991).
\newblock Sliced inverse regression for dimension reduction.
\newblock \textit{Journal of the American Statistical Association}
  86(414):316--327.

\bibitem[{Li(1992)}]{Li1992PHD}
Li, K.C. (1992).
\newblock On principal hessian directions for data visualization and dimension
  reduction: Another application of stein's lemma.
\newblock \textit{Journal of the American Statistical Association}
  87(420):1025--1039.

\bibitem[{Mercer(1909)}]{mercer1909functions}
Mercer, J. (1909).
\newblock Functions of positive and negative type and their commection with the
  theory of integral equations.
\newblock \textit{Philosophical Transactions of the Royal Society London}
  209:415--416.

\bibitem[{Sch{\"o}lkopf et~al.(1997)Sch{\"o}lkopf, Smola and
  M{\"u}ller}]{scholkopf1997kernel}
Sch{\"o}lkopf, B., Smola, A. and M{\"u}ller, K.R. (1997).
\newblock Kernel principal component analysis.
\newblock In: \textit{International conference on artificial neural networks},
  pp. 583--588. Springer.

\bibitem[{Weng et~al.(2003)Weng, Zhang and Hwang}]{weng2003candid}
Weng, J., Zhang, Y. and Hwang, W.S. (2003).
\newblock Candid covariance-free incremental principal component analysis.
\newblock \textit{IEEE Transactions on Pattern Analysis and Machine
  Intelligence} 25(8):1034--1040.

\bibitem[{Wu(2008)}]{WuHM}
Wu, H.M. (2008).
\newblock Kernel sliced inverse regression with applications to classification.
\newblock \textit{Journal of Computational and Graphical Statistics - J COMPUT
  GRAPH STAT} 17:590--610.

\bibitem[{Wu et~al.(2013)Wu, Liang and Mukherjee}]{Wu}
Wu, Q., Liang, F. and Mukherjee, S. (2013).
\newblock Kernel sliced inverse regression: Regularization and consistency.
\newblock \textit{Abstract and Applied Analysis} 2013:Article ID 540725, 11
  pages.

\bibitem[{Xu et~al.(2022)Xu, Cui and Cheng}]{cheng2021online}
Xu, J., Cui, W. and Cheng, H. (2022).
\newblock Online sparse sliced inverse regression for high dimensional
  streaming data.
\newblock \textit{International Journal of Wavelets, Multiresolution and
  Information Processing} .

\bibitem[{Yeh et~al.(2009)Yeh, Huang and Lee}]{Yeh}
Yeh, Y.R., Huang, S.Y. and Lee, Y.J. (2009).
\newblock Nonlinear dimension reduction with kernel sliced inverse regression.
\newblock \textit{IEEE Trans. Knowl. Data Eng.} 21:1590--1603.

\bibitem[{Zhang and Wu(2019)}]{zhang2019online}
Zhang, N. and Wu, Q. (2019).
\newblock Online learning for supervised dimension reduction.
\newblock \textit{Mathematical Foundations of Computing} 2(2):95.

\end{thebibliography}


\end{document}